\documentclass[%
preprint,
 amsmath,amssymb,
 aps,
pra,
]{revtex4-2}

\usepackage{graphicx}% Include figure files
\usepackage{dcolumn}% Align table columns on decimal point
\usepackage{bm}% bold math
\usepackage{amsmath}
\newcommand{\ket}[1]{\mbox{$|{ #1}\rangle$}}
\newcommand{\bra}[1]{\mbox{$\langle { #1}|$}}
\usepackage{xcolor}

\begin{document}

%\preprint{APS/123-QED}%\usepackage{graphicx} % Required for inserting images

%\usepackage{tikz}
%\usetikzlibrary{quantikz}

\title{Clauser-Horne-Shimony-Holt Bell-inequality Violability 
with the Full Poincar\'e-Bloch Sphere}

\author{Carlos Cardoso-Isidoro}
\author{Enrique J. Galvez}
 \email{egalvez@colgate.edu}
\affiliation{Department of Physics and Astronomy, Colgate University, 13 Oak Drive, Hamilton, NY 13346, U.S.A.}%Kiko Galvez }
\date{\today}
%\section{Abstract}
\begin{abstract}
    Linearly polarized projections are the tacit means for performing Clauser-Horne-Shimony-Holt (CHSH) Bell-inequality tests using polarization-entangled photon pairs. The inequality is valid for all states on the Poincar\'e-Bloch sphere, but few laboratory studies have investigated violations with the full sphere. In this article, we explore the experimental verifications of the predicted violations of the CHSH inequality with Bell and non-Bell states with same and different linear and elliptically polarized basis states for each photon. We find that Bell states violate CHSH when using the same basis for both photons, regardless of their ellipticity, whereas they show {\em no violations} for photon projections in different bases. We found non-Bell maximally-entangled states for which the converse is true. 
\end{abstract}
\maketitle

\section{Introduction}
The landmark thought experiment by Einstein Podolsky and Rosen (EPR) presented the physical situation of two entangled particles described in terms of position and momentum \cite{EPR}. Bohm reframed this situation in terms of spin-1/2 particles \cite{Bohm51}. Later Bell developed a theory that provided a testable framework for quantum mechanics in the form of inequalities involving spin-1/2 particles \cite{Bell64}. A general form of this inequality applied to light was proposed by Clauser, Horne, Shimony and Holt (CHSH) \cite{CHSH}. Because polarization states form a 2-state system that is easily manipulated experimentally, the CHSH inequality has been the basis of experimental tests based on linear polarization-state projections \cite{AspectPRL82}. Entangled states based on non-separable superpositions of polarization entangled photons violate the CHSH inequality to its maximum degree, and have become a staple of this fundamental measurement, and in particular, used in landmark experiments that address various loopholes in the falsification of local-realism  \cite{WeihsPRL98,ScheidlPnas10,Rauchprl18}. The simplicity of the basis has allowed the test to be implemented as educational laboratories \cite{MitchellAJP02}.

Experimentally, CHSH tests with entangled photons have repeatedly confirmed violations using linear polarization bases \cite{weihs1998}, and more recent works have expanded Bell tests to other degrees of freedom, such as spatial modes or orbital angular momentum \cite{Leach2009, McLaren2012}. However, in these implementations, measurement settings were confined to planar subspaces of the Bloch sphere; a limited exploration of all the possible correlations. 

The space of polarization states represented by the Poincar\'e sphere, where antipode points on the sphere correspond to orthogonal states, is a parallel to the Bloch sphere describing a two-state system. Linear polarization states form only a great circle on the Poincar\'e sphere, and such a restriction overlooks the more general body of states accessible on the Bloch sphere. Allowing elliptical polarization projections enables exploration of the entire sphere, revealing more violations than those obtained with planar settings alone. Violation of inequalities with non-linear states of polarization is possible in any polarization basis formed by antipodes on the sphere. Although departing from the linear bases presents an additional experimental complication by having to make elliptical-polarization projections, making a test of the CHSH inequality with a general state of polarization is an illustrative extension of the generality of the problem. Surprisingly, there are no reports of measurements of violations of the CHSH inequality with elliptical bases. Although it is clear that the linear bases is enough, here we show that further investigation of the use of the full Bloch sphere reveals hidden subtleties.

A non-violation of the CHSH inequality with linear-state projections of a maximally-entangled non-Bell state was reported recently \cite{DavisJOSAB25}. Here we show that it can produce maximal violations by using projections in mixed bases spanning the full Poincar\'e-Bloch sphere. We investigate the violability of the CHSH inequality with maximally-entangled states, and show, for example, that there are bases where Bell states do not violate the CHSH inequality for any combinations of settings.

This article is structured as follows. In Sec.~\ref{sec:theory} we make a presentation of the conceptual rationale for the experiments we performed. It is divided into two sections: Sec.~\ref{sec:states} defines the polarization states used in this work as three families of states within the Poincar\'e sphere. Sec.~\ref{sec:correlations} explores the nonlocal correlations of measurements with these families of states, for the traditional Bell states and other maximally-entangled states. The state preparation and apparatus used for the experiment are described in Sec.~\ref{sec:app}. In Sec.~\ref{sec:Bell} we present the landscapes of possible outcomes of the CHSH parameter and our measurements sampling such spaces for several types of initial states and measurement bases. In Sec.~\ref{sec:diss} we discuss further generalizations and give concluding remarks in Sec.~\ref{sec:conc}.  
%%%%%%%%%%%%

\section{Theoretical Framework}\label{sec:theory}

The CHSH Bell test consists of considering the parameter $S$ given by \cite{CHSH}
\begin{equation}
    S=\langle AB\rangle-\langle AB'\rangle+\langle A'B\rangle+\langle A'B'\rangle,\label{eq:s}
\end{equation}
where, quantum mechanically, $\langle AB\rangle$ is the expectation value of the measurement of the state of two particles with measuring devices $A$ and $B$, where $A$ has eigenvalues $+1$ with eigenvector $\ket{a}$, and $-1$ with eigenvector 
$\ket{a_\perp}$. Similarly $B$ has eigenvalues $+1$ with eigenvector $\ket{b}$, and $-1$ with eigenvector 
$\ket{b_\perp}$.
A local-realistic hidden-variable theory must satisfy
\begin{equation}
|S|\le 2. \label{eq:s_ineq}
\end{equation}

\begin{figure}[htbp]
\centering\includegraphics[width=0.7\columnwidth]{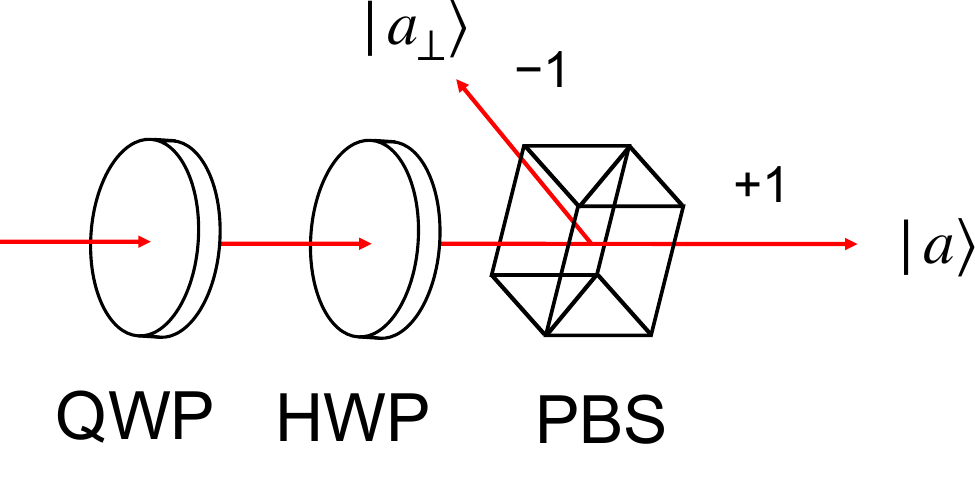}
\caption{Scheme of a splitter for any state of polarization $\ket{a}$ and the state orthogonal to it $\ket{a_\perp}$. It consists of a quarter-wave plate (QWP), half-wave plate (HWP) and a polarizing beam splitter (PBS).}
\label{fig:bs}
\end{figure}

The original version was meant to be consistent with spin-1/2 particles, but applied to photon polarization. Within such a scheme, polarization measurements mimic Stern-Gerlach measurements, where a $+1$ eigenvalue meant deflection of the spin in the ``up'' direction and $-1$ a deflection in the ``down'' direction of the Stern Gerlach apparatus. When adapted to photons with a polarizing splitter, as shown in Fig.~\ref{fig:bs}, we assign $+1$ to the photon that is transmitted (corresponding to state $\ket{a}$), and $-1$ to the photon that is reflected (in state $\ket{a_\perp}$). 

The combination of a quarter-wave plate (QWP), half-wave plate (HWP) and a fixed polarizing beam splitter can serve as a splitter of light into any state of polarization $\ket{a}$, and the state orthogonal to it, $\ket{a_\perp}$. 
This scheme involves two detectors per particle.
If only one detector is used, the polarizing beam splitter is replaced by a polarizer, and the wave plates are adjusted to transmit $\ket{a}$ and $\ket{a_\perp}$ in separate measurements.

%%%%%%%%%%%%

\subsection{General Polarization States}\label{sec:states}
For a single qubit, the measurement operator is
\begin{equation}
    \hat{A}=\ket{a}\bra{a}-\ket{a_\perp}\bra{a_\perp}.
\end{equation}
The expectation value of the measurement of a state $\psi$ with the beam splitter can be expressed in terms of probabilities
\begin{equation}
    \langle A\rangle=P(a)-P(a_\perp)
\end{equation}
or more generally in terms of the density matrix $\hat{\rho}$ of the state
\begin{equation}
    \langle A\rangle={\rm Tr}[\hat{\rho}\hat{A}].
\end{equation}

\begin{figure}[htbp]
\centering\includegraphics[width=\columnwidth]
{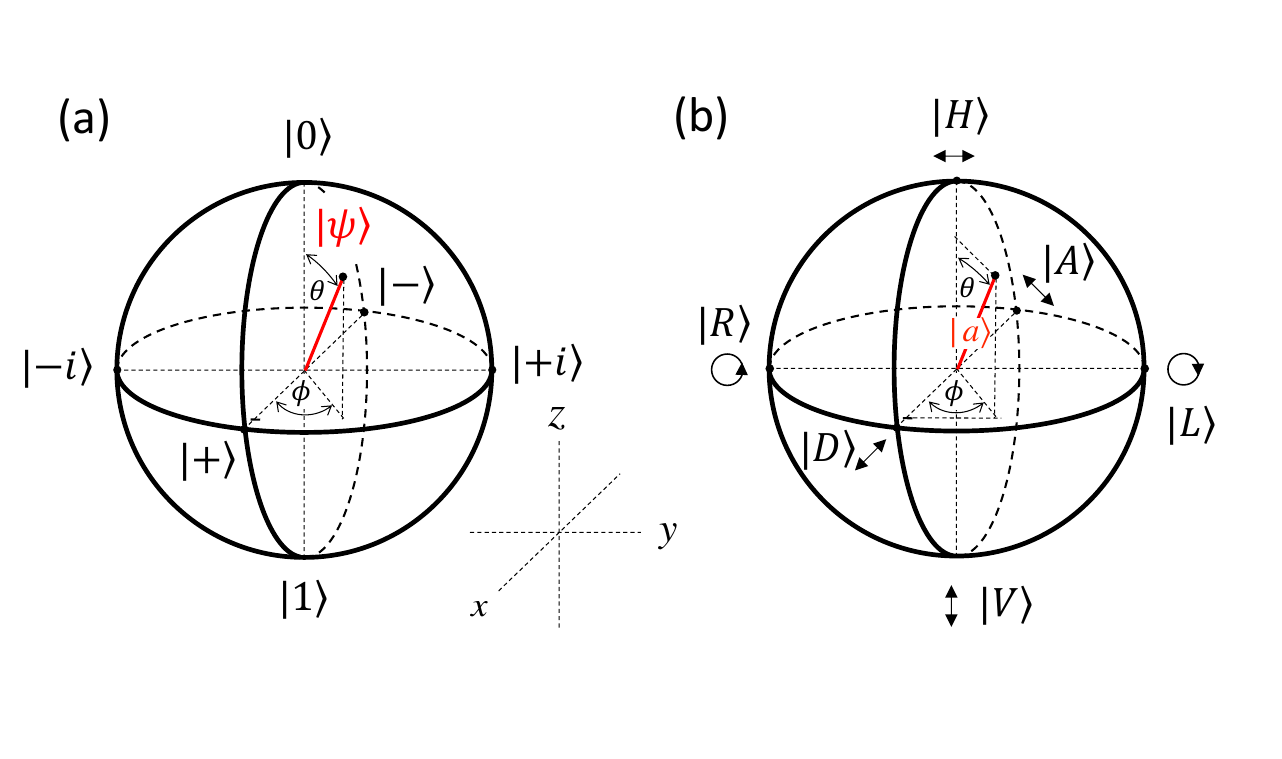}
%{FigBlockPoinc.pdf}
\caption{(a) Bloch sphere for the space of single qubits, where $\ket{\pm}=2^{-1/2}(\ket{0}\pm\ket{1})$ and $\ket{\pm i}=2^{-1/2}(\ket{0}\pm i\ket{1})$. (b) Poincar\'e-Bloch sphere, where all states of polarization are represented by a unique point on the sphere. We define the polarization states: $\ket{D}=2^{-1/2}(\ket{H}+\ket{V})$, $\ket{A}=2^{-1/2}(\ket{H}-\ket{V})$, $\ket{L}=2^{-1/2}(\ket{H}+i\ket{V})$, and $\ket{R}=2^{-1/2}(\ket{H}-i\ket{V})$.}
\label{fig:pbs}
\end{figure}

All possible states of a two-state system can be expressed as points on the Bloch sphere, where the North pole is state $\ket{0}=\begin{pmatrix}1&0\end{pmatrix}^T$ and the south pole is state $\ket{1}=\begin{pmatrix}0&1\end{pmatrix}^T$, as shown in Fig. \ref{fig:pbs}(a). These are also the states of spin-1/2 $\ket{+z}$ and $\ket{-z}$, respectively. In the case of light's polarization, all the states can be represented on the Poincar\'e sphere, shown in Fig. \ref{fig:pbs}(b). Because the state at the north pole of the Poincar\'e sphere shown is horizontal polarization ($\ket{H}$) instead of the usual right-circular ($\ket{R}$), we will call it the Poincar\'e-Bloch sphere. Using the polar coordinate system shown in the figure, a general axis on the sphere is given by 
\begin{equation}
    \vec{\rm n}=\sin\theta\cos\phi\;\hat{\rm i}+\sin\theta\sin\phi\;\hat{\rm j}+\cos\theta\;\hat{\rm k},
\end{equation} 
where $\theta\in[0,\pi]$ and $\phi\in[0,2\pi]$. The general state of polarization expressed on the horizontal-vertical basis, where $\ket{H}=\ket{0}$ and $\ket{V}=\ket{1}$, is
\begin{equation}
    \ket{a}=\begin{pmatrix}\cos(\theta/2)\\e^{i\phi}\sin(\theta/2)\end{pmatrix}.
\end{equation}
Because orthogonal states are antipodes on the sphere,
\begin{equation}
    \ket{a_\perp}=\begin{pmatrix}\sin(\theta/2)\\-e^{i\phi}\cos(\theta/2)\end{pmatrix},
\end{equation}
where we have simplified the polar angle of the antipode, $\theta\rightarrow \pi-\theta$, and the azimuthal angle, $\phi\rightarrow\phi+\pi$.
These states are eigenstates of the spin along the direction $\vec{\rm n}$, given by the operator
$\vec{\sigma}\cdot\vec{\rm n}$,
where 
$\vec{\sigma}=\hat{\sigma}_x\;\hat{\rm i}+\hat{\sigma}_y\;\hat{\rm j}+\hat{\sigma}_z\;\hat{\rm k}$, 
%$\vec{\sigma}=\hat{X}\;\hat{\rm i}+\hat{Y}\;\hat{\rm j}+\hat{Z}\;\hat{\rm k}$, 
with $\hat{\sigma}_x$, $\hat{\sigma}_y$, and $\hat{\sigma}_z$ 
%with $\hat{X}$, $\hat{Y}$, and $\hat{Z}$ 
being the Pauli matrices.

%%%%%%%%%%%%

\subsubsection{Linear hd States}
\begin{figure}[htb]
\centering\includegraphics[width=\columnwidth]{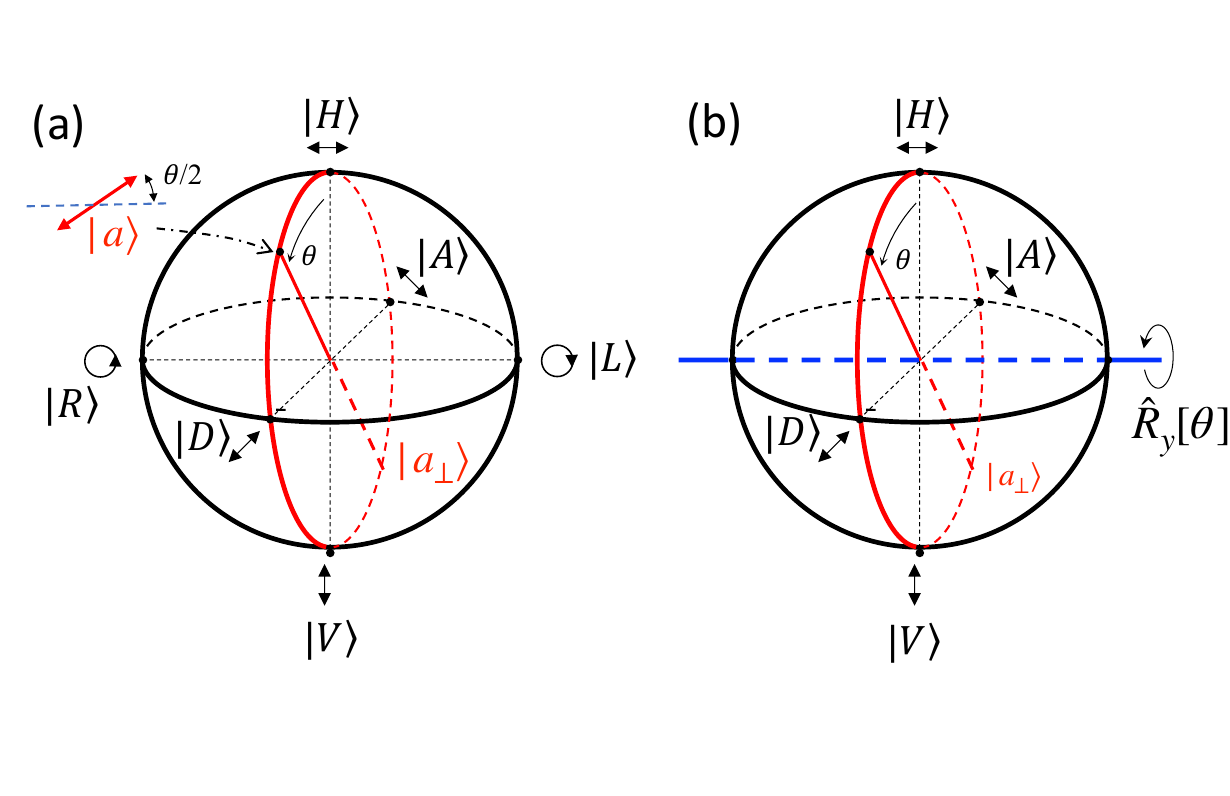}
%{Fighdstates.pdf}
\caption{(a) The locus of linear states on the Poincar\'e-Bloch sphere lie along the great circle that contains states $\ket{H}$ and $\ket{D}$, and so we call it the ``$hd$'' great circle. (b) States in this basis are reached by a rotation about the $y$ (R-L) axis.}
\label{fig:pbhd}
\end{figure}

All states of linear polarization correspond to $\phi=0$ on the sphere and are contained in the great circle that includes $\ket{H}$ and $\ket{D}$, as shown in Fig.~\ref{fig:pbhd}(a). We label these states by ``$hd$.'' A general state of linear polarization along this great circle on the Poincar\'e sphere, as shown in Fig.~\ref{fig:pbhd}(a), and oriented at an angle $\alpha=\theta/2$ is given by
\begin{equation}
    \ket{a}=\begin{pmatrix}\cos(\theta_a/2)\\\sin(\theta_a/2)\end{pmatrix}.\label{eq:ahd}
\end{equation}
The state orthogonal to it is given by
\begin{equation}
    \ket{a_\perp}=\begin{pmatrix}\sin(\theta_a/2)\\-\cos(\theta_a/2)\end{pmatrix}.\label{eq:aphd}
\end{equation}

A linear state $\ket{a}$ is detected by a single polarizer at angle $\alpha_a=\theta/2$, and state $\ket{a_\perp}$ is detected by a polarizer at an angle $\alpha_a+\pi/2$. Using the apparatus in Fig.~\ref{fig:bs} to detect the state $\ket{a}$, the QWP is set to the angle $\alpha_a$ and the HWP is set to the angle $\alpha_a/2$.

Bell-inequality violations with polarization-entangled photons are almost exclusively done by projections along the $hd$ great circle. All four Bell states can be used to measure violations with linear polarization states to the greatest degree allowed by quantum theory. This stems from the nonlocal correlations between photons in Bell states expressed in the linear bases. However, as we will see later, there are entangled states that are not correlated in the linear basis but are correlated in elliptical bases. The objective of this work is to show violations of the inequality for these states in non-linear bases.

States along this great circle are obtained by rotation about the $y$-axis, as shown in Fig.~\ref{fig:pbhd}(b). The rotation operator for vectors along the $hd$ great circle is given by \cite{NielsenChuang}
\begin{eqnarray}
    R_y[\theta]&=&e^{i\theta\; \hat{\sigma}_y/2}\\
    &=&\cos(\theta/2)\;\hat{I}-i\sin(\theta/2)\;\hat{\sigma}_y\\
%    R_y[\theta]&=&e^{i\theta\; \hat{Y}/2}\\
%    &=&\cos(\theta/2)\;\hat{I}-i\sin(\theta/2)\;\hat{Y}\\    
&=&\begin{pmatrix}\cos(\theta/2) & -\sin(\theta/2)\\ \sin(\theta/2) & \cos(\theta/2)\end{pmatrix}.
\end{eqnarray}

%%%%%%%%%%%%

\subsubsection{Elliptical $hr$ States}
\begin{figure}[htbp]
\centering\includegraphics[width=\columnwidth]{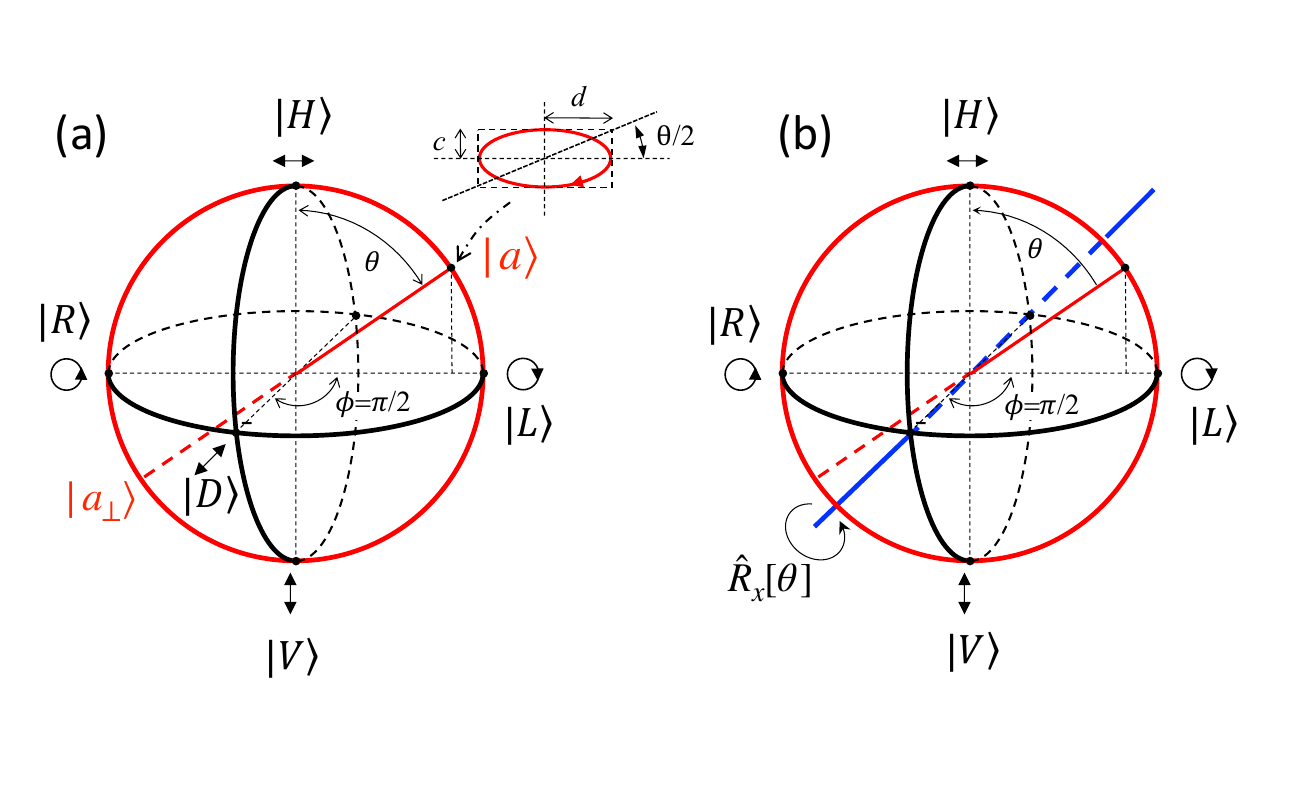}
%{Fighrstates.pdf}
\caption{(a) The locus of elliptical states on the Poincar\'e-Bloch sphere with ellipse semi-axes along the horizontal-vertical orientations. (b) States are moved along the $hr$ circle via rotations about the $x$-axis.}
\label{fig:pbp2}
\end{figure}

We wish to specify two other families of states on the Poincar\'e-Bloch sphere. One contains the elliptical states with semi-major and semi-minor axes along the Cartesian axes, as shown in Fig.~\ref{fig:pbp2}(a). A state of this kind lies along a great circle that contains states $\ket{H}$ and $\ket{R}$, as seen in the figure. We label these states as ``$hr$.'' The general state along the $hr$ great circle is represented by 
\begin{equation}
    \ket{a}=\begin{pmatrix}\cos(\theta/2)\\\pm i\sin(\theta/2)\end{pmatrix}\label{eq:ahr}
\end{equation}
with its antipode
\begin{equation}
    \ket{a_\perp}=\begin{pmatrix}\sin(\theta/2)\\\mp i\cos(\theta/2)\end{pmatrix}.\label{eq:aphr}
\end{equation}
These are states of varying ellipticity. If the length of the semi-major and semi-minor axes of the ellipse are $d$ and $c$, respectively, then the ellipticity of the polarization ellipse is 
\begin{equation}
\epsilon=\tan(\theta/2)=c/d.
\end{equation}

In general, in the reference frame where the axes are parallel to the semi-axes of the ellipse, the components of the field are out of phase by $\pi/2$. 
Thus, these states are detected with a fixed QWP with fast axis fixed and aligned with the horizontal, followed by a polarizer set to $\theta/2$ for $\ket{a}$, and $\theta/2+\pi/4$ for $\ket{a_\perp}$. Turning the polarizer selects the state along the $hr$ circle that is detected. With the apparatus of Fig.~\ref{fig:bs}, we set HWP to $\theta/4$, with preceding QWP set to 0.

The rotation operator for states along the $hr$ great circle involves rotations about the $x$-axis of the Bloch sphere, given by \cite{NielsenChuang}
\begin{eqnarray}
     \hat{R}_x[\theta]&=& e^{-i\theta\;\hat{\sigma}_x/2} \\
     &=&\cos(\theta/2)\;\hat{I}-i\sin(\theta/2)\;\hat{\sigma}_x\\
%     \hat{R}_x[\theta]&=& e^{-i\theta\;\hat{X}/2} \\
%     &=&\cos(\theta/2)\;\hat{I}-i\sin(\theta/2)\;\hat{X}\\
     &=&\begin{pmatrix} \cos(\theta/2) & -i\sin(\theta/2)\\
     -i\sin(\theta/2) & \cos(\theta/2)\end{pmatrix}.
\end{eqnarray}
This is illustrated in Fig.~\ref{fig:pbp2}(b).

%%%%%%%%%%%%

\subsubsection{Elliptical $dr$ States}
\begin{figure}[htbp]
\centering\includegraphics[width=\columnwidth]{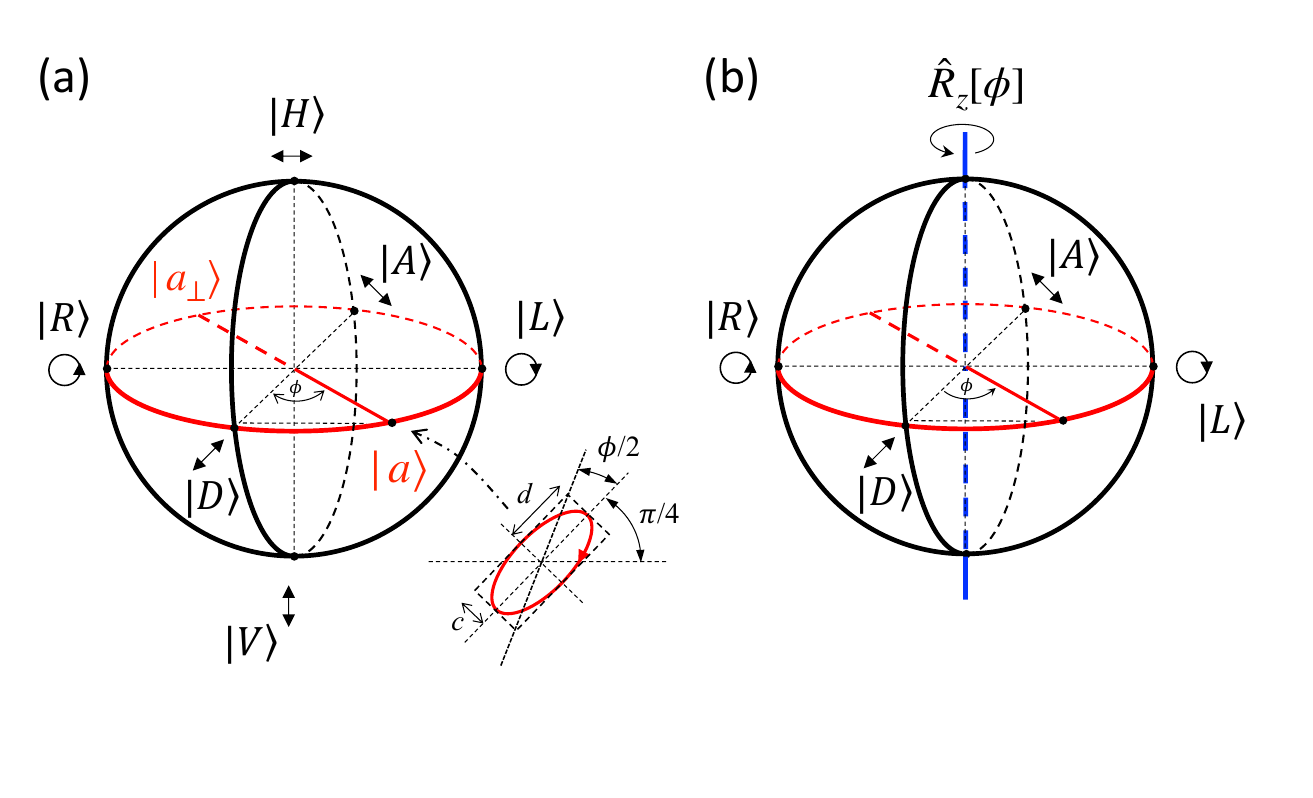}
%{Figdrstates.pdf}
\caption{(a) A locus of elliptical states on the Poincar\'e-Bloch sphere with axes along the diagonal-antidiagonal orientations. (b) states along the $dr$ great circle are transformed by a rotation about the $z$-axis, as illustrated.}
\label{fig:pbp3}
\end{figure}

A second set of polarization elliptical states lies along the great circle that contains states $\ket{D}$ and $\ket{R}$, as shown in Fig.~\ref{fig:pbp3}(a). These correspond to elliptical states with axes along the diagonals ($\pm 45^\circ$) of the Cartesian frame. We label this state ``$dr$.'' These states are
\begin{equation}
    \ket{a}=\frac{1}{\sqrt{2}}\begin{pmatrix}1\\ e^{i\phi}\end{pmatrix}\label{eq:adr}
\end{equation}
and 
\begin{equation}
    \ket{a_\perp}=\frac{1}{\sqrt{2}}\begin{pmatrix}1\\ -e^{i\phi}\end{pmatrix}.\label{eq:apdr}
\end{equation}
These are states of varying ellipticity. If the semi-major and semi-minor axes of the ellipse are $d$ and $c$ respectively, then the ellipticity is 
\begin{equation}
    \epsilon=\tan(\phi/2)=c/d.
\end{equation}
These states are detected with a QWP with fast axis fixed along the diagonal ($45^\circ$ to the horizontal), followed by a polarizer at angle $\phi/2+\pi/4$ for state $\ket{a}$, and $\phi/2+3\pi/4$ for state $\ket{a_\perp}$. Turning the polarizer selects the state along the $dr$ great circle that is detected.

Rotations along the $hd$ great circle are made by rotations about the $z$-axis. The operator is given by \cite{NielsenChuang}
\begin{eqnarray}
    \hat{R}_z&=&e^{i\phi\;\hat{\sigma}_z/2}\\
    &=&\cos(\phi/2)\;\hat{I}-i\sin(\phi/2)\;\hat{\sigma}_z\\
%    \hat{R}_z&=&e^{i\phi\;\hat{Z}/2}\\
 %   &=&\cos(\phi/2)\;\hat{I}-i\sin(\phi/2)\;\hat{Z}\\
    &=&\begin{pmatrix}e^{-i\phi/2} & 0 \\0 & e^{i\phi/2}\end{pmatrix}.
\end{eqnarray}

%%%%%%%%%%%%

\subsection{Correlation Parameter}\label{sec:correlations}
The operator for measuring the first qubit in state $\ket{a}$ and the second qubit in state $\ket{b}$ is
\begin{equation}
    \hat{A}\otimes\hat{B}=\left(\ket{a}\bra{a}-\ket{a_\perp}\bra{a_\perp}\right)\otimes\left(\ket{b}\bra{b}-\ket{b_\perp}\bra{b_\perp})\right).\label{eq:corrp}
\end{equation}
The expectation value used in the Bell test
is
\begin{equation}
    \langle AB\rangle=P(a,b)-P(a,b_\perp)-P(a_\perp,b)+P(a_\perp,b_\perp)\label{eq:pab},
\end{equation}
or equivalently,
\begin{equation}
    \langle AB\rangle={\rm Tr}[(\hat{A}\otimes\hat{B})\hat{\rho}].\label{eq:atb}
\end{equation}
A convenient expression useful for calculations is \cite{NielsenChuang,Gamel2016}
\begin{equation}
    \langle AB\rangle=\hat{a}^T\hat{T}\hat{b},\label{eq:atb}
\end{equation}
where $\hat{a}$ and $\hat{b}$ are real vectors on the Poincar\'e-Bloch sphere corresponding to the measurement directions. $\hat{T}$ is a real $3\times3$ ``correlation'' matrix with elements given by%\cite{horodecki1995}?
\begin{equation}
    T_{ij}={\rm Tr}[\rho(\sigma_i\otimes\sigma_j)].\label{eq:tij}
\end{equation}
%%%%%%%%%%%%

\subsubsection{Linear $hd$ States}
If $\ket{a}$ and $\ket{b}$ are linear states (i.e., lying along the great circle $hd$ of Fig.~\ref{fig:pbhd}(a)) at angles $\alpha_a$ and $\alpha_b$ relative to the horizontal, respectively,
the probability of measuring the Bell state 
\begin{equation}
    \ket{\Phi^+}=\frac{1}{\sqrt{2}}\left(\ket{H}\ket{H}+\ket{V}\ket{V}\right)
\end{equation}
in state $\ket{a}\ket{b}$ using Eqs. \ref{eq:ahd} and \ref{eq:aphd} is
\begin{equation}
    P(a,b)=\frac{1}{2}\cos^2(\alpha_a-\alpha_b).\label{eq:phi+}  
\end{equation}
Combining the probabilities for the other combinations involving $\ket{a_\perp}$ and $\ket{b_\perp}$, we get
\begin{equation}
    \langle AB\rangle_{hd-\Phi^+}=\cos2(\alpha_a-\alpha_b).\label{eq:hdphip}
\end{equation}
This relation is also 
%When $\ket{a}$ and $\ket{b}$ are linear states (i.e., along the $hd$ great circle), then 
\begin{equation}
    \langle AB\rangle_{hd}\equiv E(\alpha_a,\alpha_b),
\end{equation}
where $E(\alpha_a,\alpha_b)$ is the correlation parameter used in the standard CHSH Bell test with linear states proposed by Aspect et al \cite{AspectPRL82}.

Let us now consider another entangled state:
\begin{equation}
    \ket{\Phi^{'+}}=\frac{1}{\sqrt{2}}\left(\ket{H}\ket{H}+i\ket{V}\ket{V}\right).\label{eq:phipp}
\end{equation}
For this state, the correlation parameter is
\begin{equation}
    \langle AB\rangle_{hd-\Phi'^+}=\cos2\alpha_a\cos2\alpha_b,\label{eq:hdphipp}
\end{equation}

We were initially motivated by the work of Ref.~\cite{DavisJOSAB25}, which uses the state
\begin{equation}
    \ket{\Psi'^+}=\frac{1}{\sqrt{2}}\left(\ket{H}\ket{V}+i\ket{V}\ket{H}\right)
\end{equation}
followed by a quarter-wave plate with fast axis at $45^\circ$. This state is given by
\begin{equation}
	\ket{\chi}=\frac{1}{2}\left(\ket{H}\ket{H}+i\ket{H}\ket{V}-\ket{V}\ket{H}+i\ket{V}\ket{V}\right),\label{eq:psipjc}
\end{equation}
which can also be expressed as
\begin{equation}
    \ket{\chi}=\frac{1}{\sqrt{2}}\left(\ket{H}\ket{L}-\ket{V}\ket{R}\right).\label{eq:chi}
\end{equation}
For this state, the correlation parameter is
\begin{equation}
    \langle AB\rangle_{hd-\chi}=-\sin2\alpha_a\cos2\alpha_b.\label{eq:hdpsipjc}
\end{equation}

Notice that the last two correlation parameters, Eqs.~\ref{eq:hdphipp} and \ref{eq:hdpsipjc},  involve no angle correlations. That is, measuring the outcome at one angle is independent of its relation to the other angle. 

\subsubsection{Elliptical $hr$ States}

If we use the $hr$ basis states, Eqs.~\ref{eq:ahr} and \ref{eq:aphr}, we get for the correlation of the three mentioned states:
\begin{eqnarray}
    \langle AB\rangle_{hr-\Phi^+}&=&\cos(\theta_a+\theta_b)\label{eq:hrphip}\\
      \langle AB\rangle_{hr-\Phi'^+}&=&\cos\theta_a\cos\theta_b\label{eq:hrphipp}\\
          	\langle AB\rangle_{hr-\chi}&=&\cos(\theta_a)\sin(\theta_b)\label{eq:hrpsippjc}.
\end{eqnarray}
Notice that the outcomes show a correlation between angles only for $\ket{\Phi^+}$. 

\subsubsection{Elliptical $dr$ States}

If we use the $dr$ basis states of Eqs.~\ref{eq:adr} and \ref{eq:apdr}, the correlation parameter for the  mentioned states is:
\begin{eqnarray}
    \langle AB\rangle_{dr-\Phi^+}&=&\cos(\phi_a+\phi_b)\label{eq:drphip}\\
      \langle AB\rangle_{dr-\Phi'^+}&=&\sin(\phi_a+\phi_b)\label{eq:drphipp}\\
    \langle AB\rangle_{dr-\chi}&=&\sin(\phi_a)\cos(\phi_b)\label{eq:drpsippjc}.
\end{eqnarray}
States $\ket{\Phi^+}$ and $\ket{\Phi'^+}$ %and $\ket{\Psi'^+}$ 
have angle correlations, but not $\ket{\chi}$.

\subsubsection{Linear-Elliptical $hdhr$ States}
Because the states of the two photons occupy distinct Hilbert spaces, we can use different bases for each. So, let us consider using the $hd$ basis for photon 1 and the $hr$ basis for photon 2. Working out the correlation parameter for the mentioned states gives
\begin{eqnarray}
    \langle AB\rangle_{hdhr-\Phi'^+}&=&\cos2(\alpha_a-\theta_b/2)\label{eq:hdhrphipp},\\
    	\langle AB\rangle_{hdhr-\chi}&=&-\sin2(\alpha_a-\theta_b/2)\label{eq:hdhrchi}.
\end{eqnarray}
Note that now there is a correlation in the angles for state $\ket{\chi}$.

Conversely, and surprisingly for us, in this mixed basis state $\ket{\Phi^+}$ does not show correlations:
\begin{equation}
    \langle AB\rangle_{hdhr-\Phi^+}=\cos2\alpha_a\cos\theta_b\label{eq:phiphdhr}.
\end{equation}
A similar outcome is found for the other three Bell states.

\subsection{Maximum CHSH violation}
We have seen that given an entangled state and measurement basis, we find angle correlations in the correlation parameter (Eqs.~\ref{eq:corrp}, \ref{eq:pab} and \ref{eq:atb}) for certain bases. Horodecki {\em et al.} found a criterion for the maximum value of $S$ to be given by
\cite{horodecki1995}
\begin{equation}
    S_{\rm max} = 2\sqrt{M(\rho)},
\end{equation}
where $M(\rho)$ is the sum of the two largest eigenvalues of $\hat{T}^T\hat{T}$, with $T$ as defined by Eq.~\ref{eq:tij}, which depends on the density matrix. 
States $\ket{\Phi^+}$, $\ket{\Phi^{\prime+}}$, and $\ket{\chi}$, all yield a maximum value of $S$ given by the Tsirelson bound: $S_{\rm max}=2\sqrt{2}$.
However, this criterion does not specify the bases where the maximum value of $S$ is attained. Given the bases considered so far, we can do a search to find them. 

\begin{figure}[h]
    \centering
    \includegraphics[width=\columnwidth]{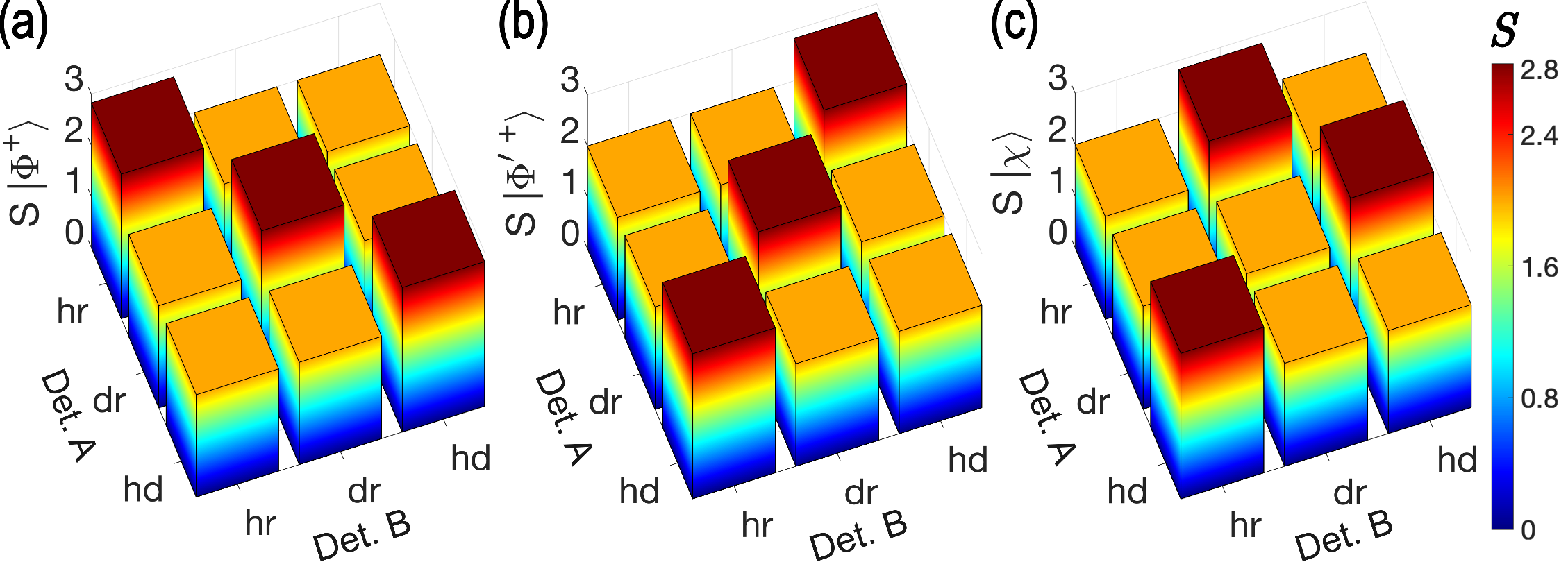}
    \caption{Bar graph of maximal value of $S$ reachable for  states $\ket{\Phi^+}$ (a), $\ket{\Phi'^+}$ (b) and $\ket{\chi}$ (c), for different basis combinations for each photon detection. 
    }
    \label{fig:Sbasis}
\end{figure}

Fig.~\ref{fig:Sbasis} shows $S_{\rm max}$ found
for different combinations of the measurement bases,
with panels (a), (b) and (c) corresponding to states $\ket{\Phi^+}$, \ket{\Phi'^+} and $\ket{\chi}$, respectively. The results are interesting. For state $\ket{\Phi^+}$, measurements with the two bases along the same great circle on the Poincar\'e-Bloch sphere yield the maximal theoretical value $2\sqrt{2}$, while off-diagonal elements, where bases are along two different great circles, show a maximum value of 2, or no violation! Fig.~\ref{fig:Sbasis}(b) shows the corresponding measurements for $\ket{\Phi'^+}$, a diagonal shows the Tsirelson bound reached when one detector is along the basis $hd$ and the other along $hr$ or vise versa, or when both are along the $dr$ basis. 
Fig.~\ref{fig:Sbasis}(c) shows the same graph for $\ket{\chi}$, where $2\sqrt{2}$ is reached only under specific combinations of basis for each detector, but none for when the bases are along the same great circle. Notice that order also matters. These results also match the cases where angle correlations on the correlation parameter are seen or not.

\section{Experimental Apparatus}\label{sec:app}
\begin{figure}[htb]
\centering\includegraphics[width=\columnwidth]{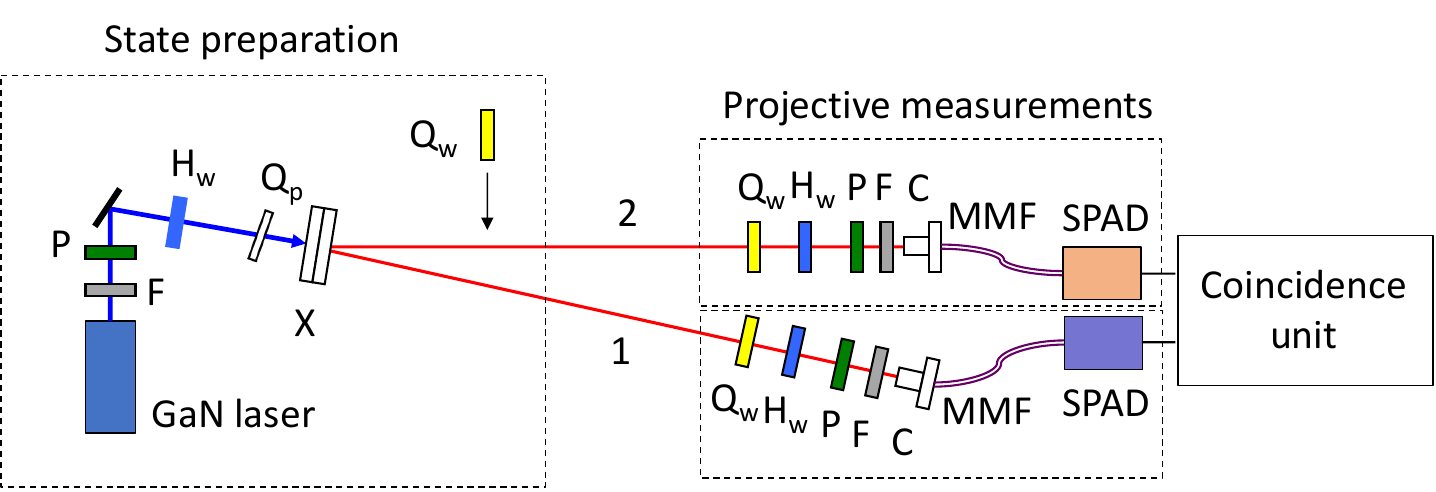}
\caption{Schematic of the apparatus. Optical components include: bandpass filters (F), polarizers (P), half-wave plates (H$_{\rm w}$), quarter-wave plates (Q$_{\rm w}$), quartz plate (Q$_p$), BBO crystals (X), collimating/mode-matching optics (C), multimode fibers (MMF), single-photon detectors (SPAD).}
\label{fig:app}
\end{figure}

We verified these predictions experimentally.
We used the apparatus shown in Fig.~\ref{fig:app}. Polarization-entangled photon pairs were produced using two crossed beta-barium-borate (BBO) crystals via Type-I spontaneous parametric down-conversion (SPDC). A polarizer, half-wave plate, and a 5-mm quartz plate were used to prepare the light of a single-mode diode laser (Toptica model Topmode-405) to prepare the state of the light. 

Bell states are normally produced in the general state
\begin{equation}
    \ket{\Phi}=\frac{1}{\sqrt{2}}\left(\ket{H}\ket{H}+e^{i\delta}\ket{V}\ket{V}\right)\label{eq:phid}
    %\\
    %,
\end{equation}
where $\delta$ is a phase that appears due to the differences in optical travel of the two polarizations through the birefringent crystal(s) in the SPDC process. The half-wave plate was adjusted to have equal absolute-value amplitudes of the two terms in the state. The phase $\delta$ was adjusted with a basis different from the HV basis. The state $\ket{\Phi^+}$ on the DA basis is:
\begin{equation}
       \ket{\Phi^+}=\frac{1}{\sqrt{2}}\left(\ket{D}\ket{D}+\ket{A}\ket{A}\right),\label{eq:phipda}
\end{equation}
so, to get $\ket{\Phi^+}$ we adjusted the tilt of the quartz plate for a minimum in coincidences when projecting the state $\ket{D}\ket{A}$ \cite{KwiatPRA99}. 

Since state $\ket{\Phi^{\prime+}}$ can also be expressed as
\begin{equation}
    \ket{\Phi^{\prime+}}=\frac{1}{\sqrt{2}}\left(\ket{D}\ket{L}+\ket{A}\ket{R}\right),\label{eq:phippda}
\end{equation}
we prepared it by
adjusting the quartz plate for a minimum in the joint signal when projecting 
the state $\ket{D}\ket{R}$ or $\ket{A}\ket{L}$. We also made state $\ket{\Phi^{\prime+}}$ by preparing state $\ket{\Phi^+}$ and adding a quarter-wave plate with fast axis horizontal in the path of photon 2.

We prepared state $\ket{\chi}$ by noticing
the relation 
\begin{equation}
    \ket{\chi}=\hat{Q}_2(-\pi/4)\ket{\Phi^{\prime+}}.
\end{equation}
We verified that we had indeed prepared the state by testing the correlations implied by Eq.~\ref{eq:chi} (projecting onto states $\ket{H}\ket{R}$ or $\ket{V}\ket{L}$).

After a state preparation section, the state was measured by elements that could project any polarization state. The elements for each photon included a quarter-wave plate, a half-wave plate, and a fixed Thompson prism-polarizer. The mounts for the waveplates were motorized so that we could take data more efficiently. After those elements we had a pair of collimating lenses in addition to a commercial fiber collimator to best mode-match the photons into multimode fibers that guide them to single-photon avalanche diode detectors (SPAD). The signal from the detectors was recorded and processed by a coincidence unit.

\section{Bell-Test Measurements}\label{sec:Bell}

In this section, we investigate the space of measurement angles that give rise to all possible values of $S$ for a given state and measurement basis.

Because the CHSH test involves four angles, without any loss of generality we can make one of them zero. For example, to detect using the $hd$ basis, we make $\alpha_a=0$. Because the expectation value $\langle AB\rangle$ also involves a measurement of state $\ket{a_\perp}$ (i.e., measuring at $\alpha_a=\pi/2$), it seems reasonable to pick an unbiased measurement between the two for the second angle: $\alpha_a^\prime=\pi/4$. Given these choices, we are left with two other angles to specify: $\alpha_b$ and $\alpha_b^\prime$. Applying this to Eqs.~\ref{eq:s} and \ref{eq:hdphip}, for $|\Phi^+\rangle$ in the {\it hd} basis, yields
\begin{equation}
    S=2\sqrt{2}\cos(\alpha_b-\alpha'_b+\pi/4)\sin(\alpha_b+\alpha'_b).\label{eq:sphiphd}
\end{equation}
The contour plot of this outcome is shown in 
\begin{figure}[htbp]
\centering\includegraphics[width=\columnwidth]{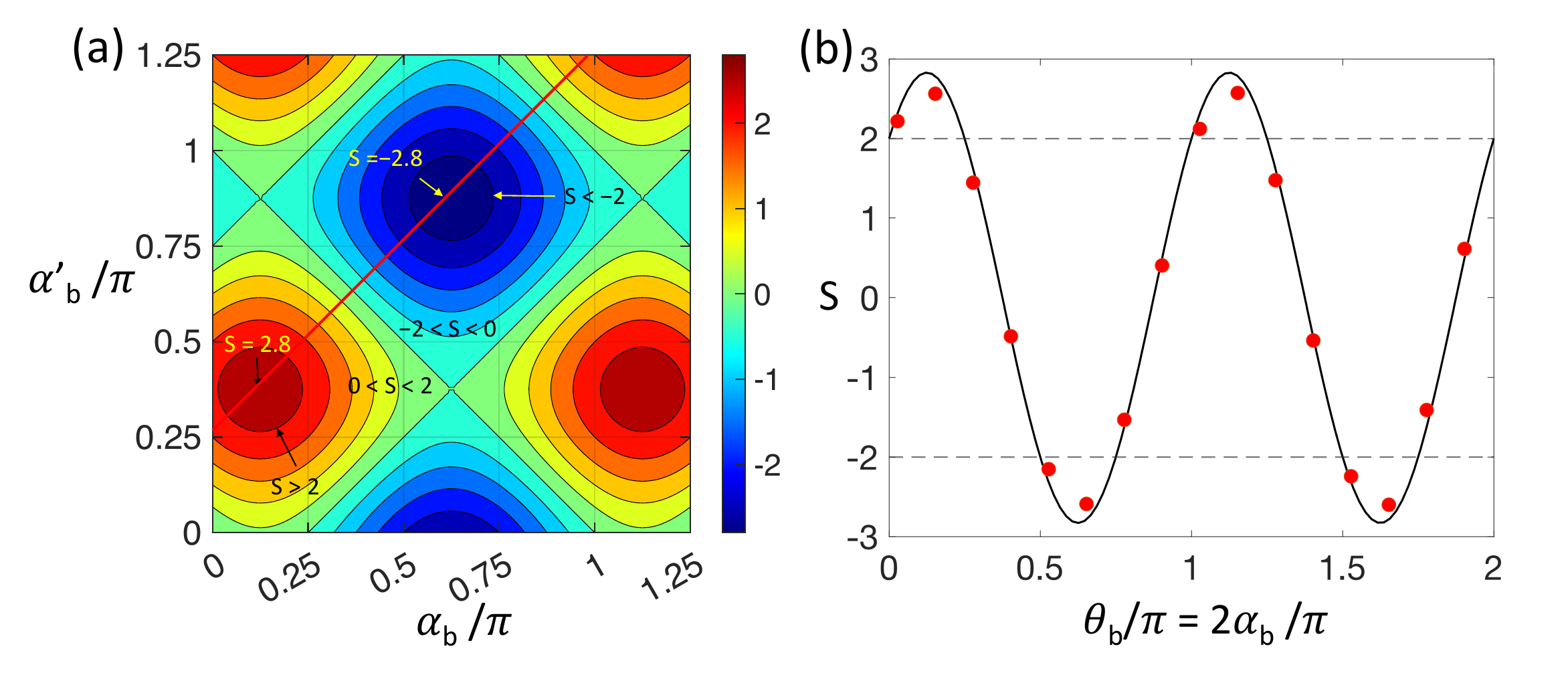}
\caption{(a) Contour map of the $S$ parameter in the CHSH Bell test of state $\ket{\Phi^+}$ in the $hd$ bases. The first and second angles are fixed to $\alpha_a=0$ and $\alpha_a'=\pi/4$, with the two other angles $\alpha_b$ and $\alpha_b'$ serving as axes. Color bar besides indicates $S$ values; (b) Graph of measurements and theory along the path shown in part (a) and Table~I, row~1.}
\label{fig:hd}
\end{figure}
Fig.~\ref{fig:hd}(a).
The axes are the possible values of $\alpha_b$ and $\alpha_b^\prime$. 
It can be seen that the angles span all possible values of $S$, which includes regions where the inequality of Eq.~\ref{eq:s_ineq} is violated (within the circular contours), and at their center the Tsirelson bound:  $|S|=2\sqrt{2} \approx 2.82$.

We tested the predictions of the graph by following a path on the contour maps where $S$ takes on all possible values along the path (e.g., following the red line in Fig.~\ref{fig:hd}(a)). The predicted value of $S$ for such a path is shown in the top entry of Table~I, and is plotted with the data in Fig.~\ref{fig:hd}(b). The data were taken at regular intervals of $\alpha_b$, with $\alpha_b'$ set according to Table~I.%\ref{tab:pathab}. 
\begin{table}\label{tab:pathab}
    \begin{center}
        \caption{Column 1: entangled states considered. Great-circle labels for the bases of projections for $a$ and $b$ are in columns 2 and 3. The angle settings for projections of $a$ are $\alpha_a=0$ and $\alpha_a'=\pi/4$ for basis $hd$, and $\phi_a=0$ and $\phi_{a'}=\pi/2$ for the $dr$ basis. Column 4 gives the relations between the angles for the angular projections $b$ for noted paths through the contours of Figs. 8-12, and column 6 gives the expected value of $S$ for each path. }
        \vspace{0.1in}
        \begin{tabular}{c|c|c|c|c}
        \hline
        State & $a$ & $b$ & Path Relation & Expected $S$  \\ \hline
        $\ket{\Phi^+}$ & $hd$ & $hd$ & $\alpha_b'=\alpha_b+\pi/4$ & $S=2\sqrt{2}\cos(2\alpha_b-\pi/4)$\\
         $\ket{\Phi'^+}$ & $hd$ & $hd$ & $\alpha_b'=\alpha_b+\pi/2$ & $S=2\sin(2\alpha_b+\pi/2)$\\
       $\ket{\Phi^+}$ & $dr$ & $dr$ & $\phi_b'=\phi_b-\pi/2$ & $S=-2\sqrt{2}\sin(\phi_b-\pi/4)$\\
        $\ket{\Phi'^+}$ & $dr$ & $dr$ & $\phi_b'=\phi_b-\pi/2$ & $S=2\sqrt{2}\cos(\phi_b-\pi/4)$\\
        $\ket{\Phi^+}$ & $hd$ & $hr$ & $\theta_b'=\theta_b+\pi$ & $S=2\sin(\theta_b+\pi/2)$\\         $\ket{\chi}$ & $hd$ & $hr$ & $\theta_b'=\theta_b+\pi/2$ & $S=2\sqrt{2}\sin(\theta_b-\pi/4)$\\ \hline
       \end{tabular}
    \end{center}
\end{table}
It can be seen that the data follows closely the expectation. The data extend into the inequality-violation zone, as predicted. Uncertainties are not shown because they were smaller than the size of the symbols.

\begin{figure}[htbp]
\centering\includegraphics[width=\columnwidth]{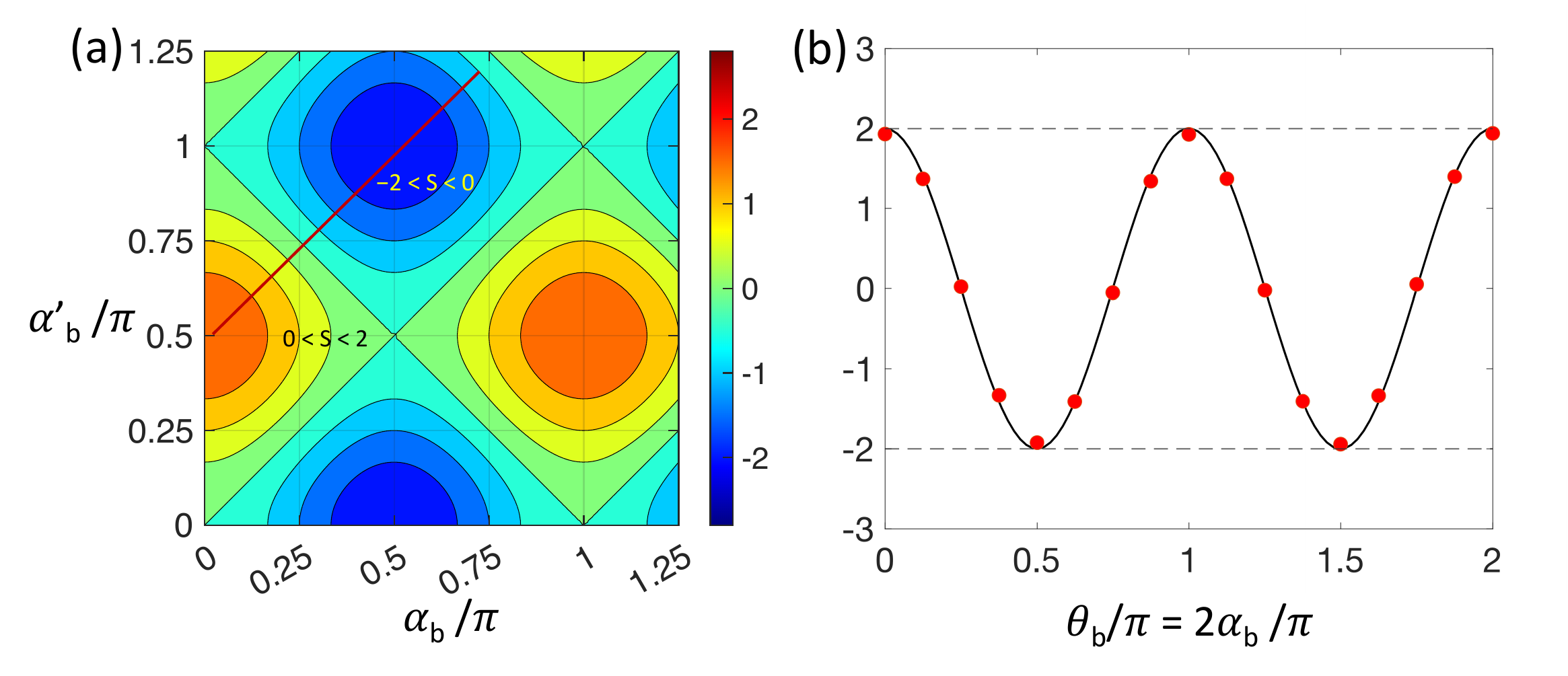}
\caption{(a) Contour map of the $S$ parameter in the CHSH Bell test of state $\ket{\Phi'^+}$ in the linear ($hd$) bases. The first and second angles are fixed to $\alpha_a=0$ and $\alpha_a'=\pi/4$, with the two other angles $\alpha_b$ and $\alpha_b'$ serving as axes. (b) Graph of measurements and theory along the path shown in part (a) and Table~I, row~2.}
\label{fig:hdpp}
\end{figure}

Applying the same angle projections for  the $S$ parameter of state $\ket{\Phi'^+}$ in the $hd$ basis yields
\begin{equation}
    S=-2\sin(\alpha_b+\alpha_b')\sin(\alpha_b-\alpha_b').\label{eq:sphipphd}
\end{equation}
The corresponding contour plot
is shown in Fig.~\ref{fig:hdpp}(a). Equations.~\ref{eq:hdphipp} and \ref{eq:hrphipp} do not show correlations for this state in the $hd$ or $hr$ bases. Correspondingly, Fig.~\ref{fig:hdpp}(a) does not show regions where the inequality is violated ($|S|>2$). This is also consistent with the predictions of Fig.~\ref{fig:Sbasis}(b). 
We took data for this case, in the same way as described for state $\ket{\Phi^+}$. Fig. \ref{fig:hdpp}(b) shows our data following a path through the maxima and minima of the contour, which confirms the expectation given in Table~I. The same situation occurs in the $hr$ basis: there are correlations for $\ket{\Phi^+}$ (Eq.~\ref{eq:hrphip}) but not for $\ket{\Phi'^+}$ (Eq.~\ref{eq:hrphipp}). Thus, the contours for those states in the $hr$ basis are similar to those in Figs.~\ref{fig:hd}(a) and \ref{fig:hdpp}(a), respectively. We also took measurements, but since they are very similar to the figures already shown, we refrain from showing the results for the sake of brevity.

\begin{figure}[htbp]
\centering\includegraphics[width=\columnwidth]{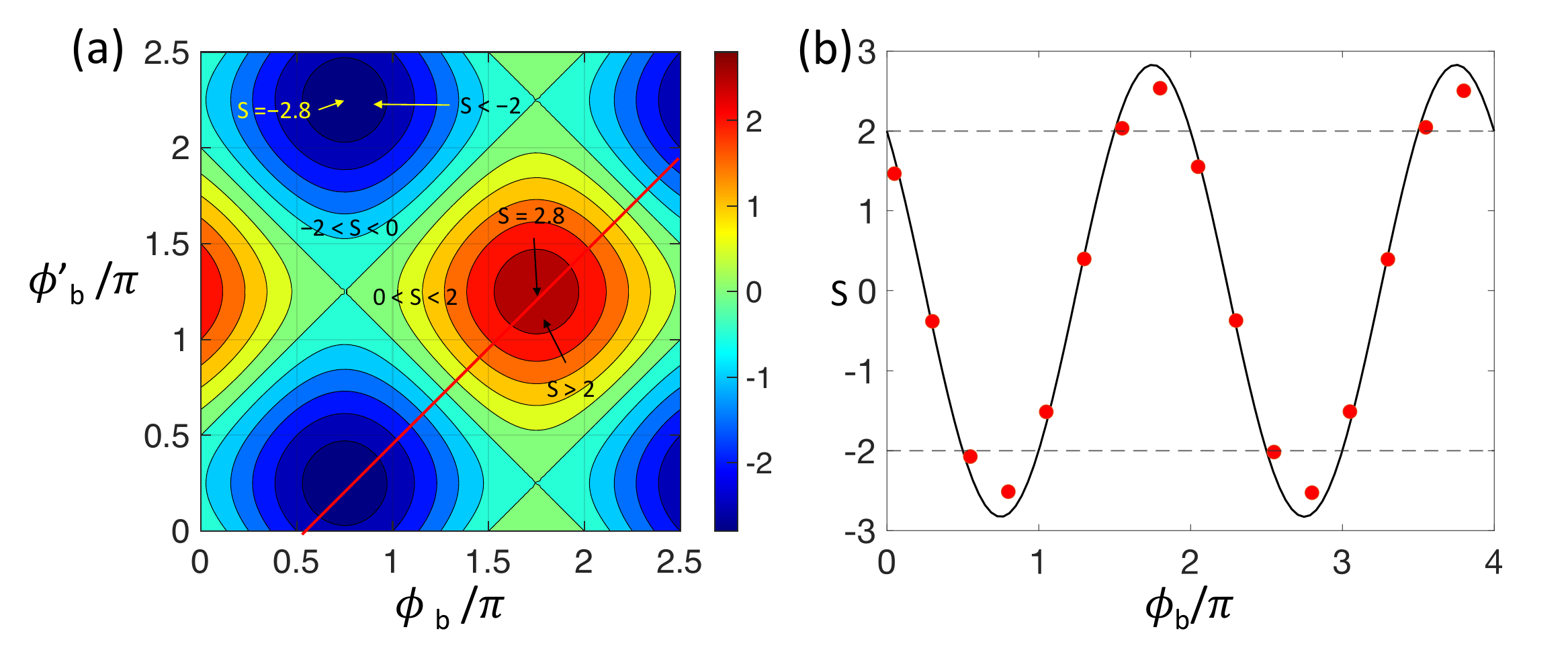}
%{Figphipdr2.pdf}
\caption{(a) Contour map of the $S$ parameter in the CHSH Bell test of state $\ket{\Phi^+}$ in the $dr$ elliptical bases. The first and second angles are fixed to $\phi_a=0$ and $\phi_a'=\pi/2$, with the two other angles $\phi_b$ and $\phi_b'$ serving as axes.  (b) Graph of measurements and theory along the path shown in part (a) and Table~I, row~3.} 
\label{fig:drpp}
\end{figure}

\begin{figure}[htbp]
\centering\includegraphics[width=\columnwidth]{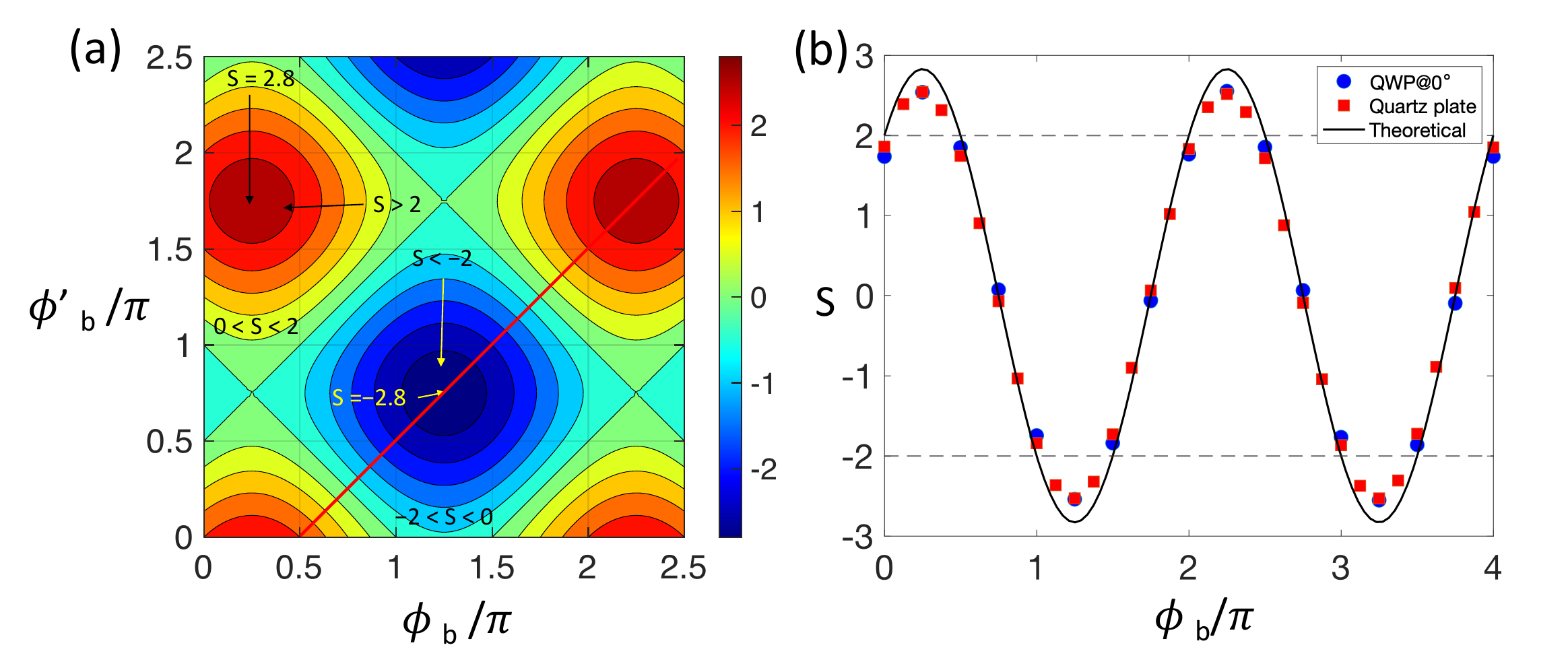}
%{Figphippdr5.pdf}
\caption{(a) Contour map of the $S$ parameter in the CHSH Bell test of state $\ket{\Phi'^+}$ in the $dr$ basis. The first and second angles are fixed to $\phi_a=0$ and $\phi_a'=\pi/2$, with the two other angles $\phi_b$ and $\phi_b'$ serving as axes.  (b) Graph of measurements and theory along the path shown in part (a) and Table~I, row~4. Two sets of measurements involve preparing the state in distinct ways. }
\label{fig:drppp}
\end{figure}

In the $dr$ basis we see correlations for both states $\ket{\Phi^+}$ and $\ket{\Phi'^+}$ via Eqs.~\ref{eq:drphip} and \ref{eq:drphipp}, respectively.  They result in 
\begin{equation}
S=-2\sqrt{2}\sin(\phi_b/2-\phi_b'/2+\pi/4)\sin(\phi_b/2+\phi_b'/2)\label{eq:sphipdr}
\end{equation}
for $\ket{\Phi^+}$, and 
\begin{equation}
    S=2\sqrt{2}\sin(\phi_b/2-\phi_b'/2+\pi/4)\cos(\phi_b/2+\phi_b'/2)\label{eq:sphippdr}
\end{equation}
for $\ket{\Phi'^+}$.
Figs.~\ref{fig:drpp} and \ref{fig:drppp} show the corresponding graphs, verifying that the inequality is violated for both states in the $dr$ basis. We note that because there were two ways of making $\ket{\Phi'^+}$, mentioned in Sec.~\ref{sec:app}, Fig.~\ref{fig:drppp}(b) shows data sets for the initial state made in these two different ways. In both cases, we see agreement with the theory, and some disagreement within the inequality-violation zone, which we attribute to a lack of perfect fidelity of our initial state.

%%%%%%%%%%%%%%%%%%%

\begin{figure}[htbp]
\centering\includegraphics[width=\columnwidth]{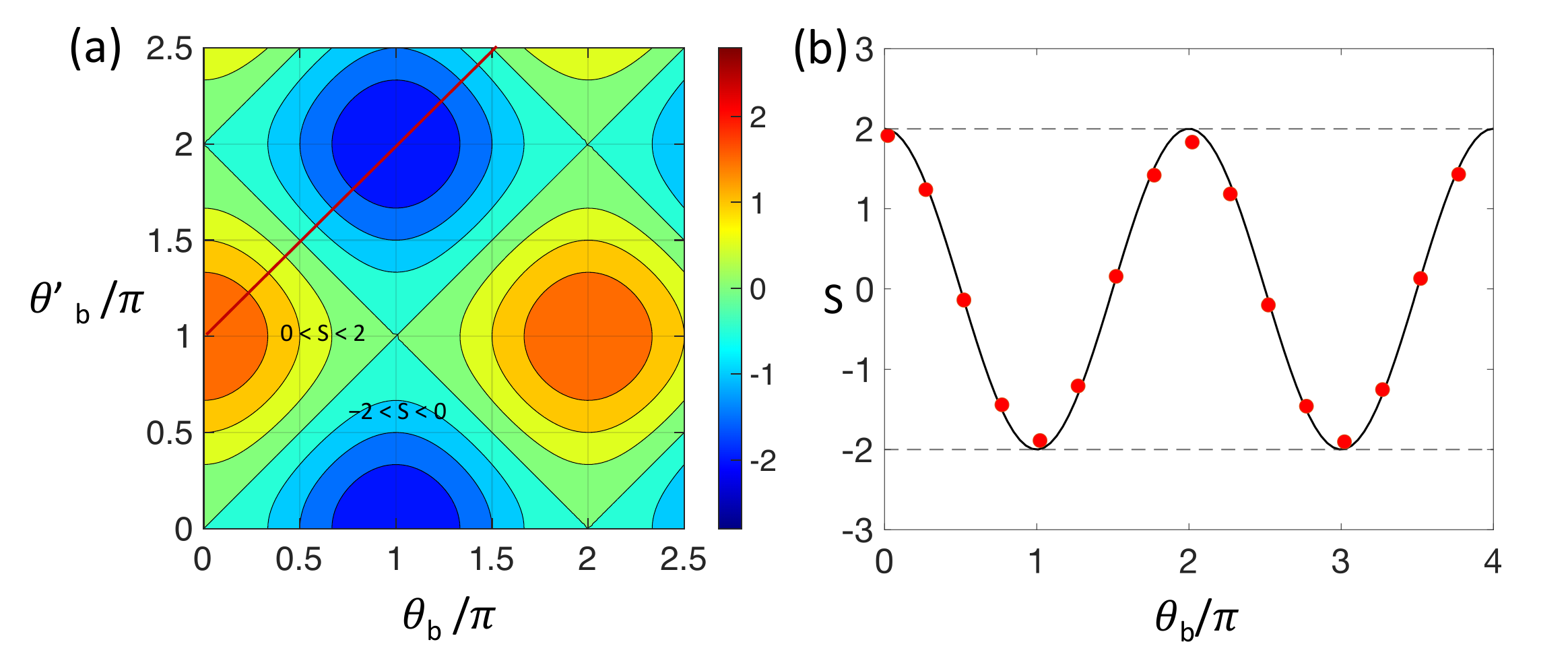}
%{Fig12Sphiphdhr.pdf}
%{Figsphiphdhr.pdf}
%\vspace{0.1in}
\caption{(a) Contour map of the $S$ parameter in the CHSH Bell test of the Bell state $\ket{\Phi^+}$ in the mixed bases, linear ($hd$) for photon 1 and elliptical-Cartesian ($hr$) for photon 2. The first and second angles are fixed to $\alpha_a=0$ and $\alpha_a'=\pi/4$, with the two other angles $\theta_b$ and $\theta_b'$ serving as axes.  (b) Graph of measurements and theory along the path shown in part (a) and Table~I, row~5.  }
\label{fig:sphiphdhr}
\end{figure}

As mentioned previously, the mixed bases presented interesting results. None of the Bell states shows correlations, and in particular, the case of $\ket{\Phi^+}$, shown in Fig.~\ref{fig:Sbasis}(a). For state $\ket{\Phi^+}$ the $S$ parameter is given by
\begin{equation}
    S=-2\sin(\theta_b/2+\theta_b'/2)\sin(\theta_b/2-\theta_b'/2),\label{eq:sphiphdhr}
\end{equation}
which is confirmed in Fig.~\ref{fig:sphiphdhr} via the predicted contour map in (a) and measurements in (b).

%%%%%%%%%%%%%%%%%%%

\begin{figure}[htbp]
\centering\includegraphics[width=\columnwidth]{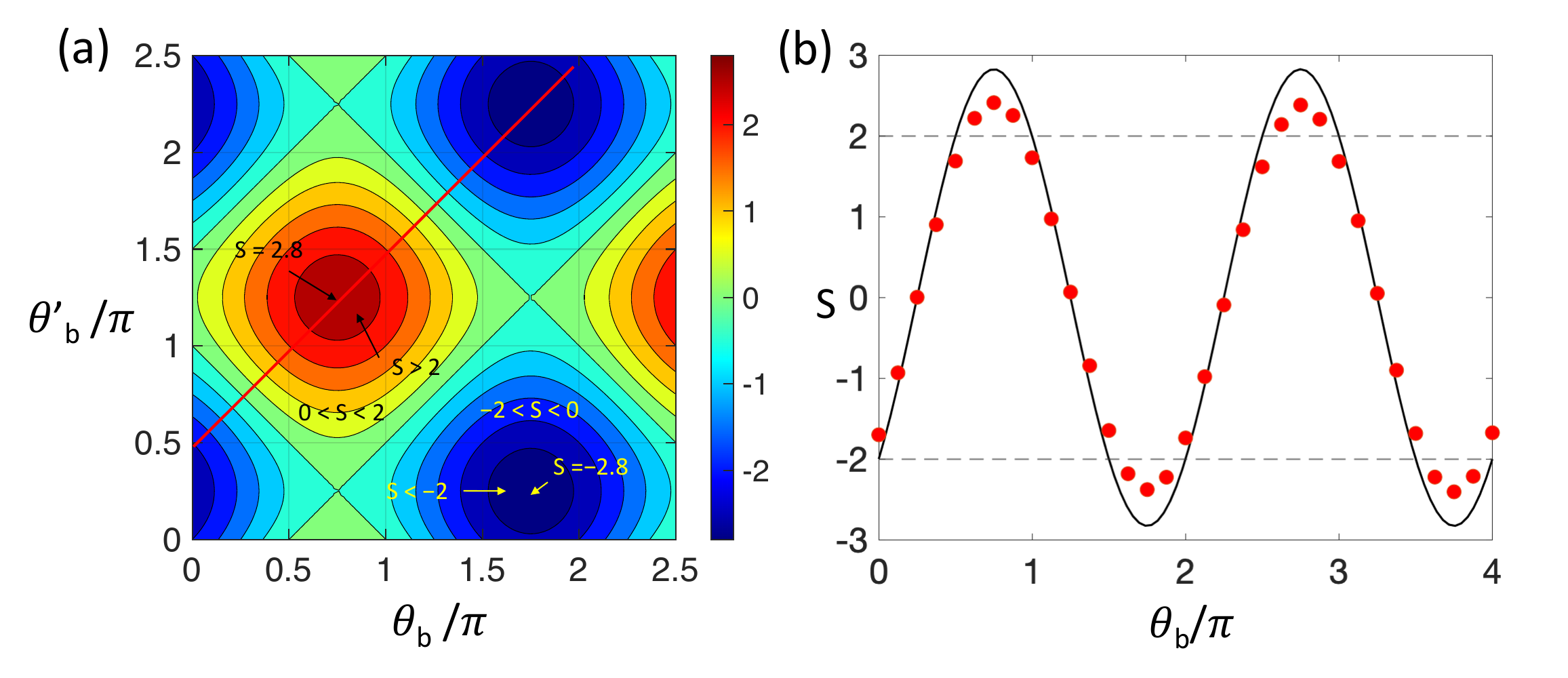}
%{Figchihdhr5.pdf}
\caption{Contour map of the $S$ parameter in the CHSH Bell test for state $\ket{\chi}$ in the mixed bases, linear ($hd$) for photon 1 and elliptical-Cartesian ($hr$) for photon 2.  The first and second angles (in $hd$ basis) are fixed to $\alpha_a=0$ and $\alpha_a'=\pi/4$, with the two other angles (in $hr$ basis) $\theta_b$ and $\theta_b'$ serving as axes. (b) Graph of measurements and theory along the path shown in part (a) and Table~I, row~6. }
\label{fig:Shdhrchi}
\end{figure}

State $\ket{\chi}$ mentioned earlier, was our original motivation. It did not have correlations in any of the three bases considered for both photons, and in these bases the inequality is not violated at any angle. For example, in the $hd$ basis $S$ is given by
\begin{equation}
S=-2\cos(\alpha_b+\alpha_b')\cos(\alpha_b-\alpha_b').\label{eq:chihd}
\end{equation}
We confirmed this experimentally, as was the case in Ref.~\cite{DavisJOSAB25}. However, 
in the mixed basis case, there are angle correlations for this state, resulting in
\begin{equation}
    S=-2\sqrt{2}\cos(\theta_b/2+\theta_b'/2)\cos(\theta_b/2-\theta_b'/2+\pi/4),\label{eq:schiphdhr}
\end{equation} 
which is confirmed in Fig.~\ref{fig:Shdhrchi}(a) where it is seen that the inequality is violated. A cut through this region verifies it, as shown in Fig.~\ref{fig:Shdhrchi}(b). We took more points for this case for a more stringent verification. We note that the maximum theoretically-predicted value of $S$ was not achieved. We attribute this to the challenge in making state $\ket{\chi}$, likely falling prey to imperfections in the setting of the angles of the waveplates.

\begin{figure}[htbp]
\centering\includegraphics[width=\columnwidth]{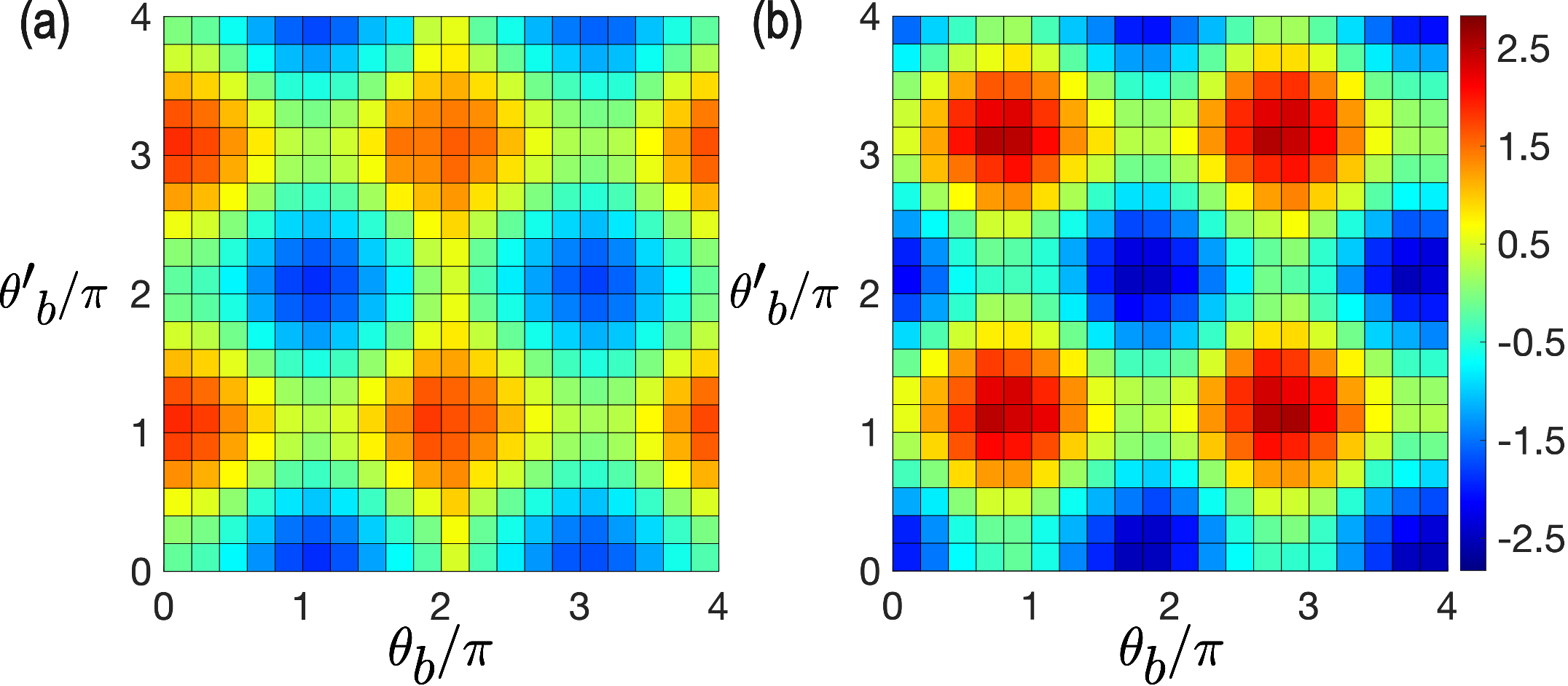}
%{FigShdhrchiphip2.pdf}
\caption{Measured contour maps of the $S$ parameter in the mixed bases, linear ($hd$) for photon 1 and elliptical-Cartesian ($hr$) for photon 2. Graph (a) corresponds to state $\ket{\Phi^+}$, whereas graph (b) to state $\ket{\chi}$. The first and second angles (in $hd$ basis) are fixed to $\alpha_a=0$ and $\alpha_a'=\pi/4$, with the two other angles (in $hr$ basis) $\theta_b$ and $\theta_b'$ serving as axes. }
\label{fig:Shdhrc}
\end{figure}

To further verify the previous results with the mixed bases, we did long scans of the angles to experimentally obtain the contour maps of states $\ket{\Phi^+}$ and $\ket{\chi}$. They are shown in Figs.~\ref{fig:Shdhrc}(a) and (b), respectively, matching the respective theoretical contours of  Figs.~\ref{fig:sphiphdhr}(a)  and \ref{fig:Shdhrchi}(a).

%%%%%%%%%%%%%%%%%%%

\section{Discussion}\label{sec:diss}
In summary, we have examined the measurements involving the CHSH Bell test for non-Bell states, which may not violate the inequality with projections of linear states of polarization. These initial states are either produced directly by the source or modified by a unitary transformation \cite{DavisJOSAB25}. One approach is to phase-transform the state into a Bell state \cite{MartinLP12}. Our approach has been to find the polarization basis of states that violate the inequality, which require the use of the full Poincar\'e-Bloch sphere. That is, making projective measurements of elliptical states. We find the conditions where this is possible and show it by making the corresponding measurements.  

We might want to ask, given an arbitrary entangled state, which measurement bases yield maximal violations? In this work we have explored experimentally only three families of bases along the three main great circles on the Poincar\'e-Bloch sphere.  Fig.~\ref{fig:Sbasis} shows predictions for the maximal values of $S$ for given states for bases along the three great circles we considered ($hd$, $hr$, $dr$).

\begin{figure}[h]
    \centering
    \includegraphics[width=\columnwidth]{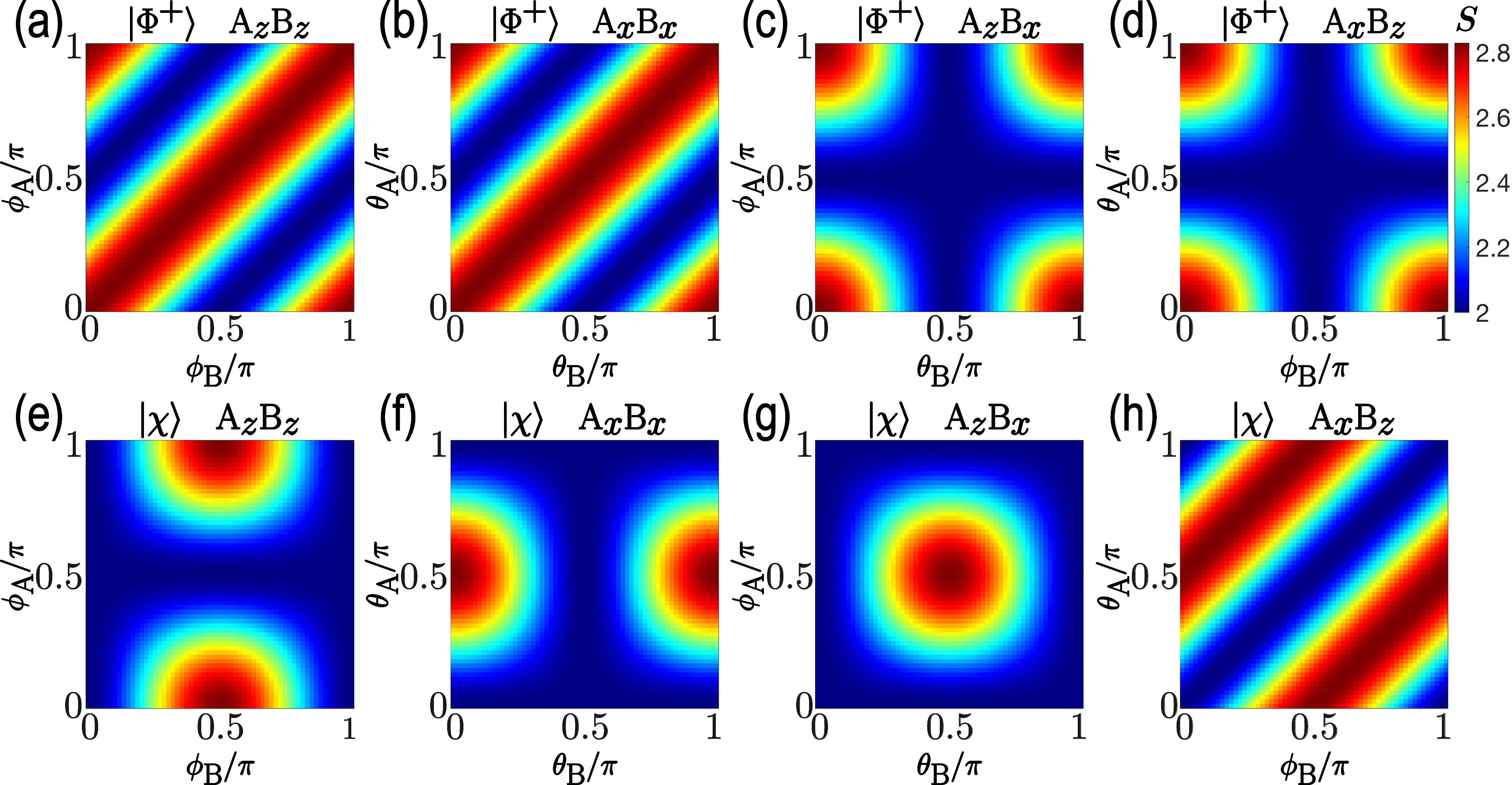}
    \caption{Maximum $S$ reachable for $\ket{\Phi^+}$ (first row) and $\ket{\chi}$ (second row) with arbitrary great circles for each detector. Measurement bases in (a,e) and (b,f) are contained in great circles on the sphere rotated about the $z$ and $x$ axes, respectively. In (c,g) one great circles rotate about $z$-axis and and the other about the $x$-axis, and the converse for (d,h). 
    }
    \label{fig:planerot}
\end{figure}

We can map the basis vectors along any pair of great circles on the Poincar\'e-Bloch sphere.
In Fig.~\ref{fig:planerot} we show maximal values of $|S|$ for bases in great circles, each rotated about an axis and both starting from the linear ($hd$) basis. For a given set of bases belonging to two great circles, we searched all possible values of the four measurement angles to find the maximum value of $|S|$. The first and second rows of the figure correspond to states $\ket{\Phi^+}$ and $\ket{\chi}$, respectively. Each column corresponds to bases along great circles at all possible angles rotated about an axis. Graphs (a) and (e) correspond to both great circles rotated about the $z$-axis with orientation specified by azimuth angles $\phi_A$ and $\phi_B$. Note that $hd$ and $hr$ bases correspond to $\phi=0$ and $\phi=\pi/2$, respectively. The graphs in panels (b) and (f) correspond to great circles rotated about the $x$-axis and orientations specified by polar angles $\theta_A$ and $\theta_B$. Also note $hd$ and $dr$ bases corresponding to $\theta=0$ and $\theta=\pi/2$, respectively. Panels (c) and (g) correspond to great circles, where one is rotated about the $z$-axis (specified by $\phi_A$), and the other, about the $x$-axis (specified by $\theta_B$). Panels (d) and (h) are the converse.

The graphs in Fig.~\ref{fig:planerot}(a) and (b), corresponding to $\ket{\Phi^+}$, highlight that, when the two bases are along the same great circle ($\phi_A=\phi_B$ or $\theta_A=\theta_B$), the Tsirelson bound is obtained. In Fig.~\ref{fig:planerot}(e) and (f), corresponding to $\ket{\chi}$, we can see non-violations for equal angles $\phi_A=\phi_B=0$ (both $hd$), $\phi_A=\phi_B=\pi/2$ (both $hr$) and $\theta_A=\theta_B=\pi/2$ (both $dr$) bases. Maximal violation is seen for $\phi_A=0$, $\phi_B=\pi/2$ ($hd$-$hr$), but not the converse, $\phi_A=\pi/2$, $\phi_B=0$ ($hr$-$hd$).  For bases along great circles rotated about orthogonal axes, seen in  \ref{fig:planerot}(c) and (d), $\ket{\Phi^+}$ shows no violations except approaching $\theta=0$ and $\phi=0$, which corresponds to the two circles being the same ($hd$). For intermediate orientations, the violation is reduced to $2<|S_{\rm max}|<2.82$. The regions of violation of $\ket{\chi}$, shown in \ref{fig:planerot}(g) and (h), are interesting. The graphs demonstrate that, in order to achieve a maximal violation and due to the states' symmetry, the bases' great circles have to be the same for state $\ket{\Phi^+}$, regardless of the great circles' absolute orientation on the sphere; and in certain great circles contained in orthogonal planes for state $\ket{\chi}$.

The previous plots emphasize that the CHSH violation depends both on the state and on the relative orientation of the bases' great circles. All confirm the Horodecki criterion.  
From a different perspective, if a non-Bell state was obtained via a unitary transformation of a Bell state, which achieves the maximum Tsirelson bound, then such a state should also exhibit nonlocal correlations and maximal violation in the basis of the adjoint of the states that show nonlocal correlations with Bell states \cite{GisinPLA91}. Our results confirm that any maximally entangled states can reach the Tsirelson bound with a suitable choice of measurement bases \cite{Scarani}.

If one detector's basis is within a given great circle on the Poincar\'e-Bloch sphere, then the choice of basis for the other detector can be obtained by searching the possible bases until the best value of $S$ is obtained.

Although the present protocol allows access to the full Poincaré-Bloch sphere, the existence of a CHSH violation depends on the intrinsic strength of quantum correlations in the shared state, as quantified by the Horodecki criterion \cite{horodecki1995}. States that are weakly entangled may remain local even under the most general measurement settings, however, the same experimental data taken with our approach in a general setup can still reveal nonclassicality through a steering test \cite{cavalcanti2009,girdhar2016} or detect nonseparability by means of an entanglement witness \cite{lewenstein2000, altepeter2005}.

\section{Conclusion}\label{sec:conc}
We have introduced a generalized CHSH test in which both detectors perform independent elliptical projective measurements, spanning the full Poincaré-Bloch sphere. With this approach, the violation of the CHSH inequality constitutes a strong certification of nonclassicality. By expanding the linear measurement bases to elliptical bases, our model can reveal CHSH violations in regions of state space that linear-state tests would miss, compensate for phase errors or misalignments introduced by imperfections in optical elements and, therefore, identify the optimal measurement orientations for maximizing $S$. This is beneficial in realistic optimal experiments, where nonideal calibrations can degrade the observed violation under standard bases.

Because CHSH nonlocality implies steering and entanglement, when a Bell violation is observed with our scheme, it will automatically follow steering and entanglement certification. However, if a violation is not observed, one can still apply steering inequalities or witnesses on selected subsets of bases, thus having a framework that can encompass witness or steering studies as sub-cases. Thus, while our scheme provides a device-independent test at the strongest level (Bell nonlocality), it also offers a unified platform for detecting weaker forms of quantum correlation.

The mapping of the violability landscape over the full Poincaré-Bloch sphere with light's polarization presented in this study provides new insights to the geometric understanding of quantum nonlocality and Bell tests. It can help uncover regions of the Bloch sphere where standard CHSH test of a given entangled state fails with basis states with real coefficients (linear polarization for the case of light) but succeeds with basis states with complex coefficients (elliptical polarization). This approach strengthens the rigor of CHSH tests and opens new paths of exploration and application of quantum nonlocality.
%\newpage

\section{Acknowledgements} E.G. thanks A. Forbes for kind support during a sabbatical leave at U. of the Witwatersrand, where this project was conceived. We thank F. De Zela for useful input in the genesis of this project, as well as J. Davis and C. Jackman for motivating this study. This work was supported by National Science Foundation grant PHY-2409587.
\bibliographystyle{unsrt}
\bibliography{BlochPoincare} 

@article{Rauchprl18,
	author = {Rauch, D. and Handsteiner, J. and Hochrainer, A. and Gallicchio, J. and Friedman, A. S. and Leung, Ca. and Liu, B. and Bulla, L. and Ecker, S. and Steinlechner, F. and Ursin, R. and Hu, B. and Leon, D. and Benn, C. and Ghedina, A. and Cecconi, Ma. and Guth, A.H. and Kaiser, D.I. and Scheidl, T. and Zeilinger, A.},
	doi = {10.1103/PhysRevLett.121.080403},
	issue = {8},
	journal = {Phys. Rev. Lett.},
	month = {Aug},
	numpages = {9},
	pages = {080403},
	publisher = {American Physical Society},
	title = {Cosmic {B}ell Test Using Random Measurement Settings from High-Redshift Quasars},
	url = {https://link.aps.org/doi/10.1103/PhysRevLett.121.080403},
	volume = {121},
	year = {2018}}

@article{WeihsPRL98,
	author = {Weihs, G. and Jennewein, T. and Simon, C. and Weinfurter, H. and Zeilinger, A.},
	doi = {10.1103/PhysRevLett.81.5039},
	issue = {23},
	journal = {Phys. Rev. Lett.},
	numpages = {0},
	pages = {5039--5043},
	publisher = {American Physical Society},
	title = {Violation of {B}ell's Inequality under Strict {E}instein Locality Conditions},
	url = {https://link.aps.org/doi/10.1103/PhysRevLett.81.5039},
	volume = {81},
	year = {1998}}

@article{ScheidlPnas10,
	author = {T. Scheidl and R. Ursin and J. Kofler and S. Ramelow and X.-S. Ma and T. Herbst and L. Ratschbacher and A. Fedrizzi and N.K. Langford and T. Jennewein and Anton Z.},
	doi = {10.1073/pnas.1002780107},
	eprint = {https://www.pnas.org/doi/pdf/10.1073/pnas.1002780107},
	journal = {PNAS},
	number = {46},
	pages = {19708-19713},
	title = {Violation of local realism with freedom of choice},
	url = {https://www.pnas.org/doi/abs/10.1073/pnas.1002780107},
	volume = {107},
	year = {2010}}

@article{KwiatPRA99,
	author = {Kwiat, P.G. and Waks, E. and White, A. G. and Appelbaum, I. and Eberhard, P.H.},
	doi = {10.1103/PhysRevA.60.R773},
	issue = {2},
	journal = {Phys. Rev. A},
	numpages = {0},
	pages = {R773--R776},
	publisher = {American Physical Society},
	title = {Ultrabright source of polarization-entangled photons},
	volume = {60},
	year = {1999}}

@article{MartinLP12,
	author = {Martin, A. and Smirr, J. -L. and Kaiser, F. and Diamanti, E. and Issautier, A. and Alibart, O. and Frey, R. and Zaquine, I. and Tanzilli, S.},
	date = {2012/06/01},
	doi = {10.1134/S1054660X12060060},
	id = {Martin2012},
	isbn = {1555-6611},
	journal = {Laser Phys.},
	number = {6},
	pages = {1105--1112},
	title = {Analysis of elliptically polarized maximally entangled states for {B}ell inequality tests},
	url = {https://doi.org/10.1134/S1054660X12060060},
	volume = {22},
	year = {2012}}

@article{GisinPLA91,
	author = {N. Gisin},
	doi = {https://doi.org/10.1016/0375-9601(91)90805-I},
	issn = {0375-9601},
	journal = {Phys. Lett. A},
	number = {5},
	pages = {201-202},
	title = {Bell's inequality holds for all non-product states},
	url = {https://www.sciencedirect.com/science/article/pii/037596019190805I},
	volume = {154},
	year = {1991}}

@book{NielsenChuang,
	author = {M.A. Nielsen and I.L. Chuang},
	publisher = {Cambridge University Press},
	title = {Quantum computation and quantum information},
	year = {2010}}

@article{AspectPRL82,
	author = {Aspect, A. and Dalibard, J. and Roger, G.},
	doi = {10.1103/PhysRevLett.49.1804},
	issue = {25},
	journal = {Phys. Rev. Lett.},
	month = {Dec},
	numpages = {0},
	pages = {1804--1807},
	publisher = {American Physical Society},
	title = {Experimental Test of {B}ell's Inequalities Using Time-Varying Analyzers},
	url = {https://link.aps.org/doi/10.1103/PhysRevLett.49.1804},
	volume = {49},
	year = {1982}}

@article{DavisJOSAB25,
	author = {J.J.J. Davis and C.L. Jackman and R. Leonhardt and P.J. Werbos and M.D. Hoogerland},
	doi = {10.1364/JOSAB.552007},
	journal = {J. Opt. Soc. Am. B},
	number = {6},
	pages = {1227--1235},
	publisher = {Optica Publishing Group},
	title = {Phase-shifted {B}ell states},
	url = {https://opg.optica.org/josab/abstract.cfm?URI=josab-42-6-1227},
	volume = {42},
	year = {2025}}

@article{MitchellAJP02,
	author = {Dehlinger, D. and Mitchell, M. W.},
	doi = {10.1119/1.1498860},
	eprint = {https://pubs.aip.org/aapt/ajp/article-pdf/70/9/903/7530140/903\_1\_online.pdf},
	issn = {0002-9505},
	journal = {Am. J.  Phys.},
	number = {9},
	pages = {903-910},
	title = {Entangled photons, nonlocality, and {B}ell inequalities in the undergraduate laboratory},
	url = {https://doi.org/10.1119/1.1498860},
	volume = {70},
	year = {2002}}

@article{chsh,
	author = {Clauser, J.F. and Horne, M.A. and Shimony, A. and Holt, R.A.},
	doi = {10.1103/PhysRevLett.23.880},
	issue = {15},
	journal = {Phys. Rev. Lett.},
	numpages = {0},
	pages = {880--884},
	publisher = {American Physical Society},
	title = {Proposed Experiment to Test Local Hidden-Variable Theories},
	url = {https://link.aps.org/doi/10.1103/PhysRevLett.23.880},
	volume = {23},
	year = {1969}}

@book{Bohm51,
	author = {D. Bohm},
	publisher = {Prentice-Hall, Englewood Cliffs},
	title = {Quantum Theory},
	year = {1951}}

@book{Scarani,
author = {V. Scarani},
title = {Bell Nonlocality},
publisher = {Oxford U. P., Oxford},
year = {2019}
}

@article{Bell64,
	author = {Bell, J. S.},
	doi = {10.1103/PhysicsPhysiqueFizika.1.195},
	issue = {3},
	journal = {Phys. Phys. Fiz.},
	numpages = {6},
	pages = {195--200},
	publisher = {American Physical Society},
	title = {On the {E}instein {P}odolsky {R}osen paradox},
	volume = {1},
	year = {1964}}

@article{EPR,
	author = {Einstein, A. and Podolsky, B. and Rosen, N.},
	doi = {10.1103/PhysRev.47.777},
	issue = {10},
	journal = {Phys. Rev.},
	numpages = {0},
	pages = {777--780},
	publisher = {American Physical Society},
	title = {Can Quantum-Mechanical Description of Physical Reality Be Considered Complete?},
	volume = {47},
	year = {1935}}

@article{girdhar2016,
  title = {All two-qubit states that are steerable via {C}lauser-{H}orne-{S}himony-{H}olt-type correlations are {B}ell nonlocal},
  author = {Girdhar, P. and Cavalcanti, E.G.},
  journal = {Phys. Rev. A},
  volume = {94},
  issue = {3},
  pages = {032317},
  numpages = {8},
  year = {2016},
  publisher = {American Physical Society},
  doi = {10.1103/PhysRevA.94.032317},
  url = {https://link.aps.org/doi/10.1103/PhysRevA.94.032317}
}

@article{lewenstein2000,
  title = {Optimization of entanglement witnesses},
  author = {Lewenstein, M. and Kraus, B. and Cirac, J. I. and Horodecki, P.},
  journal = {Phys. Rev. A},
  volume = {62},
  issue = {5},
  pages = {052310},
  numpages = {16},
  year = {2000},
  publisher = {American Physical Society},
  doi = {10.1103/PhysRevA.62.052310},
  url = {https://link.aps.org/doi/10.1103/PhysRevA.62.052310}
}

@article{altepeter2005,
  title = {Experimental Methods for Detecting Entanglement},
  author = {Altepeter, J. B. and Jeffrey, E. R. and Kwiat, P. G. and Tanzilli, S. and Gisin, N. and Ac\'{\i}n, A.},
  journal = {Phys. Rev. Lett.},
  volume = {95},
  issue = {3},
  pages = {033601},
  numpages = {4},
  year = {2005},
  publisher = {American Physical Society},
  doi = {10.1103/PhysRevLett.95.033601},
  url = {https://link.aps.org/doi/10.1103/PhysRevLett.95.033601}
}

@article{horodecki1995,
title = {Violating {B}ell inequality by mixed spin-12 states: necessary and sufficient condition},
journal = {Phys. Lett. A},
volume = {200},
number = {5},
pages = {340-344},
year = {1995},
issn = {0375-9601},
doi = {https://doi.org/10.1016/0375-9601(95)00214-N},
url = {https://www.sciencedirect.com/science/article/pii/037596019500214N},
author = {R. Horodecki and P. Horodecki and M. Horodecki}
}

@article{cavalcanti2009,
  title = {Experimental criteria for steering and the {E}instein-{P}odolsky-{R}osen paradox},
  author = {Cavalcanti, E. G. and Jones, S. J. and Wiseman, H. M. and Reid, M. D.},
  journal = {Phys. Rev. A},
  volume = {80},
  issue = {3},
  pages = {032112},
  numpages = {16},
  year = {2009},
  publisher = {American Physical Society},
  doi = {10.1103/PhysRevA.80.032112},
  url = {https://link.aps.org/doi/10.1103/PhysRevA.80.032112}
}

@article{weihs1998,
  title = {Violation of {B}ell's Inequality under Strict {E}instein Locality Conditions},
  author = {Weihs, G. and Jennewein, T. and Simon, C. and Weinfurter, H. and Zeilinger, A.},
  journal = {Phys. Rev. Lett.},
  volume = {81},
  issue = {23},
  pages = {5039--5043},
  numpages = {0},
  year = {1998},
  publisher = {American Physical Society},
  doi = {10.1103/PhysRevLett.81.5039},
  url = {https://link.aps.org/doi/10.1103/PhysRevLett.81.5039}
}

@article{Leach2009,
author = {J. Leach and B. Jack and J. Romero and M. Ritsch-Marte and R.W. Boyd and A.K. Jha and S.M. Barnett and S. Franke-Arnold and M. J. Padgett},
journal = {Opt. Express},
number = {10},
pages = {8287--8293},
publisher = {Optica Publishing Group},
title = {Violation of a {B}ell inequality in two-dimensional orbital angular momentum state-spaces},
volume = {17},
year = {2009},
url = {https://opg.optica.org/oe/abstract.cfm?URI=oe-17-10-8287},
doi = {10.1364/OE.17.008287},
}

@article{McLaren2012,
author = {M. McLaren and M. Agnew and J. Leach and F.S. Roux and M.J. Padgett and R.W. Boyd and A. Forbes},
journal = {Opt. Express},
number = {21},
pages = {23589--23597},
publisher = {Optica Publishing Group},
title = {Entangled {B}essel-{G}aussian beams},
volume = {20},
year = {2012},
url = {https://opg.optica.org/oe/abstract.cfm?URI=oe-20-21-23589},
doi = {10.1364/OE.20.023589},
}

@article{Gamel2016,
  title = {Entangled Bloch spheres: Bloch matrix and two-qubit state space},
  author = {Gamel, O.},
  journal = {Phys. Rev. A},
  volume = {93},
  issue = {6},
  pages = {062320},
  numpages = {18},
  year = {2016},
  month = {Jun},
  publisher = {American Physical Society},
  doi = {10.1103/PhysRevA.93.062320},
  url = {https://link.aps.org/doi/10.1103/PhysRevA.93.062320}
}

\end{document}